\documentclass[runningheads]{llncs}
\usepackage{bdlfoe}
\usepackage[hidelinks]{hyperref}
\hypersetup{colorlinks=true,linkcolor=blue,citecolor=blue,urlcolor=blue}
\usepackage{orcidlink}
\usepackage{version}

\newcommand{\doi}[1]{{doi:\href{http://doi.org/#1}{\nolinkurl{#1}}}}
\renewcommand{\url}[1]{{\href{#1}{\nolinkurl{#1}}}}

\spnewtheorem{sdef}{Definition}{\bfseries}{\rmfamily}
\spnewtheorem{sass}{Assumption}{\bfseries}{\rmfamily}
\spnewtheorem{snot}{Notation}{\bfseries}{\rmfamily}

\titlerunning
 {On an Ordinary Expansion of First-Order Belnap-Dunn Logic}
\authorrunning
 {C.A. Middelburg}

\begin{document}

\title{On an Ordinary Expansion of \\ First-Order Belnap-Dunn Logic}
\author{C.~A. Middelburg\,\orcidlink{0000-0002-8725-0197}}
\institute{Informatics Institute, Faculty of Science, University of
           Amsterdam, \\
           Science Park~900, 1098~XH Amsterdam, the Netherlands \\
           \href{mailto:C.A.Middelburg@uva.nl}{C.A.Middelburg@uva.nl}}

\maketitle

\begin{abstract}
This paper concerns an expansion of first-order Belnap-Dunn logic whose 
connectives and quantifiers all have a counterpart in classical logic.
The language and logical consequence relation of this paradefinite logic 
are defined, a sequent calculus proof system for this logic is 
presented, and the soundness and completeness of this proof system is 
established.
It is shown that the defined logic distinguishes itself from the many 
other paradefinite logics that are usually considered equally classical 
by the classical laws of logical equivalence that hold for it.
It is further argued that the defined logic is the most natural 
paradefinite logic relative to the version of classical logic with the 
same language.
Moreover, a simple embedding of the defined logic in that version of 
classical logic is presented and the potential of the logic for dealing 
with inconsistencies and incompletenesses in inductive machine learning 
is discussed.
\begin{keywords} 
Belnap-Dunn logic, sequent calculus, natural paradefinite logic, 
embedding, concept learning
\end{keywords}
\begin{acm-classcode} 
F.4.1, I.2.6
\end{acm-classcode}
\begin{msc-classcode} 
03B50 (Primary)\,\, 03B53 (Secondary) 
\end{msc-classcode}
\end{abstract}

\section{Introduction}
\label{INTRO}

This paper draws attention to an expansion of first-order Belnap-Dunn 
logic~\cite{AB63a} (also knows as the logic of First-Degree Entailment)
whose language is the same as the language of a version of classical 
logic.
In the expansion in question, which is called \foBDif, the added 
connectives are a falsity connective and an implication connective for 
which the standard deduction theorem holds.
Various expansions of propositional and first-order Belnap-Dunn logic
have been studied earlier, but most of them are not as closely related 
to classical logic as \foBDif\ simply because the added connectives have 
no counterpart in classical logic.
Examples of such expansions are BD$\mathrm{\Delta}$~\cite{SO14a},
F4CC~\cite{KZ20a}, and QLET$_\mathrm{F}$~\cite{ARCC22a}.  
In many cases, the added connectives can be defined in terms of the
connectives of \foBDif\ (see details in~\cite{Mid24b}).
\foBDif\ is dubbed a ``conventional expansion'' of first-order 
Belnap-Dunn logic in the title of this paper because the added 
connectives have a counterpart in classical logic.

\foBDif\ is a paradefinite logic.
This means that it is both a paraconsistent logic, i.e.\ a logic in 
which not every formula is a logical consequence of each two formulas of 
which one is the negation of the other, and a paracomplete logic, i.e.\ 
a logic in which not, for each two formulas of which one is the negation 
of the other, one or the other is a logical consequence of every set of 
formulas.
These properties make \foBDif\ interesting: it can serve as the 
underlying logic of theories that are inconsistent and/or incomplete.

In this paper, the language and logical consequence relation of \foBDif\ 
are rigorously defined, a sequent calculus proof system for \foBDif\ is 
presented, and the soundness and completeness of the presented proof 
system are established.
For the version of first-order classical logic with the same language as 
\foBDif, a sound and complete sequent calculus proof system can be 
obtained by adding two inference rules for the negation connective to 
the presented sequent calculus proof system of \foBDif.
This suggests that the logical consequence relations of \foBDif\ and the 
version of classical logic with the same language are closely related.

To illustrate the classical nature of the connectives and quantifiers of 
\foBDif, a number of classical laws of logical equivalence are given 
which distinguishes the logical equivalence relation of \foBDif\ from 
the logical equivalence relation of the many other logics that are 
usually considered equally classical.
On the basis of the way in which the logical consequence relation of 
\foBDif\ can be obtained from the logical consequence relation of the 
version of classical logic with the same language, it is further argued 
that \foBDif\ is the most natural paradefinite logic with respect to 
that version of classical logic.

Because of the classical nature of the connectives and quantifiers of 
\foBDif, there exists a simple embedding of $\foBDif$ into the version 
of classical logic with the same language as $\foBDif$.
The embedding concerned is also presented.
In addition, the potential of \foBDif\ for dealing with inconsistencies 
and incompletenesses in inductive machine learning is briefly 
discussed.
The discussion points out that the given embedding may be practically 
relevant.

The propositional fragment of \foBDif\ has been discussed in several 
earlier papers, including~\cite{AA96a,AA98a,AA17a,Pyn99a}, but without 
exception quite casually.

In the field of paraconsistent and paracomplete logics, there is no full
agreement on what the term ``first-order Belnap-Dunn logic'' stands for.
For example, one time it is a logic with equality and another time it is 
not, one time function symbols of positive arity are included and 
another time they are excluded, and one time the logical consequence 
relation is a relation on sets of formulas that may contain free 
variables and another time it is a relation on sets of formulas that may 
not contain free variables.
For this reason, we describe the expansion of first-order Belnap-Dunn 
logic this paper is concerned with clearly and in full detail.

For those readers who may wonder whether the differences between 
\foBDif\ and the other expansions of BD mentioned above 
(BD$\mathrm{\Delta}$, F4CC, and QLET$_\mathrm{F}$) matter, 
a summary of the main similarities and dissimilarities between \foBDif\ 
and these other expansions of BD is given in the concluding remarks of
this paper.
Some of the similarities and dissimilarities mentioned require a brief 
look at the cited papers on the logics in question, others require a 
very careful reading of those papers, and still others are the results 
of the research reported in~\cite{Mid24b}.

The structure of this paper is as follows.
First, the language and logical consequence relation of \foBDif\ are 
defined (Sections~\ref{LANGUAGE} and~\ref{CONSEQUENCE}). 
Next, a sequent calculus proof system for \foBDif\ is presented
(Section~\ref{PROOF-SYSTEM}).
After that, the classical nature of the connectives and quantifiers 
of \foBDif\ is illustrated by means of classical laws of logical 
equivalence that hold for it (Section~\ref{EQUIVALENCE}) and it is 
argued that \foBDif\ is the most natural paradefinite logic relative to 
the version of classical logic with the same language (Section~\ref{NATURAL}).
Following this, the embedding of \foBDif\ into classical logic is 
presented (Section~\ref{EMBEDDING}) and the potential relevance of 
\foBDif\ to inductive machine learning is discussed 
(Section~\ref{MACHINE-LEARNING}). 
Finally, some concluding remarks are made (Section~\ref{CONCLUSIONS}).
The proofs of some of the presented theorems are given in appendices.

Because we will almost exclusively talk about first-order logics with 
equality, we will mostly leave out the qualifications ``first-order'' 
and ``with equality'' in the rest of this paper.

Previous versions of this paper provide both a fairly comprehensive 
overview of \foBDif\ and a study of the interdefinability 
of \foBDif\ with other expansions of first-order Belnap-Dunn logic.
The current version only provides a revision of the overview of 
\foBDif.
Another paper, to wit~\cite{Mid24b}, provides a revision of the 
interdefinability study.
The paper has been split into two to improve both parts.

\section{The Language of \foBDif}
\label{LANGUAGE}

First the alphabet of \foBDif\ is introduced and then the terms and 
formulas of \foBDif\ are defined for a fixed but arbitrary alphabet. 

\subsection{Alphabet}

The language of \foBDif\ is based on a number of assumptions concerning 
the symbols used. 
These assumptions must be made explicit before the alphabet of \foBDif\
can be introduced.

\begin{sass}
\label{ass-symbols}
It is assumed that the following sets of symbols have been given:
\begin{itemize}
\item
a countably infinite set $\SVar$ of \emph{variables};
\item
for each $n \in \Nat$, a countable set $\Func{n}$ of 
\emph{function symbols of arity $n$};
\item
for each $n \in \Nat$, a countable set $\Pred{n}$ of 
\emph{predicate symbols of arity $n$}.
\end{itemize}
It is also assumed that all these sets are mutually disjoint and 
disjoint from the set 
$\set{\False,\Not,\CAnd,\COr,\IImpl,\forall,\exists}$ 
and that ${=} \in \Pred{2}$.
\end{sass}
Each choice of the assumed sets of symbols gives rise to a different 
instance of~\foBDif.
\begin{sdef}
\label{def-alphabet}
\sloppy
The \emph{logical symbols} of \foBDif\ are the symbols from the set
$\set{\False,\Not,\CAnd,\COr,\IImpl,\forall,\exists}$.
The \emph{non-logical symbols} of \foBDif\ are the symbols from the set
$\Union \set{\Func{n} \where n \in \Nat} \union
 \Union \set{\Pred{n} \where n \in \Nat}$.
The \emph{alphabet} of the language of $\foBDif$ consists of
the logical symbols of \foBDif,
the non-logical symbols of \foBDif, and
the variables from $\SVar$.
\end{sdef}
The function symbols of arity $0$ are also known as 
\emph{constant symbols} and
the predicate symbols of arity $0$ are also known as 
\emph{proposition symbols}.

\subsection{Terms and formulas}
The language of $\foBDif$ consists of formulas.
The formulas of $\foBDif$ are constructed from the symbols in the 
alphabet of the language of $\foBDif$ according to the formation rules 
given below.
\begin{sdef}
\label{def-terms}
The set of all \emph{terms} of $\foBDif$, 
written $\STerm$, is inductively defined by the following 
formation rules:
\begin{enumerate}
\item
if $x \in \SVar$, then $x \in \STerm$;
\item
if $c \in \Func{0}$, then $c \in \STerm$;
\item
if $f \in \Func{n+1}$ and 
$t_1,\ldots,t_{n+1} \in \STerm$, then 
$f(t\sb1,\ldots,t_{n+1}) \in \STerm$.
\end{enumerate}
The set of all \emph{closed terms} of $\foBDif$ is the subset of 
$\STerm$ that can be formed by applying formation rules~2 and~3 
only.
\end{sdef}

\begin{sdef}
\label{def-formulas}
The set of all \emph{formulas} of $\foBDif$, 
written $\SForm$, is inductively defined by the following 
formation rules:
\begin{enumerate}
\item
if $p \in \Pred{0}$, then 
$p \in \SForm$;
\item
if $P \in \Pred{n+1}$ and 
$t_1,\ldots,t_{n+1} \in \STerm$, then 
$P(t_1,\ldots,t_{n+1}) \in \SForm$;
\item
$\False \in \SForm$;
\item
if $A \in \SForm$, then $\Not A \in \SForm$;
\item
if $A_1,A_2 \in \SForm$, then 
$A_1 \CAnd A_2,\, A_1 \COr A_2,\, A_1 \IImpl A_2 \in \SForm$;
\item
if $x \in \SVar$ and $A \in \SForm$, then 
$\CForall{x}{A},\, \CExists{x}{A} \in \SForm$.
\end{enumerate}
The set $\SAForm$ of all \emph{atomic formulas} of $\foBDif$ is the 
subset of $\SForm$ that can be formed by applying formation rules~1 
and~2 only.
\end{sdef}

\begin{snot}
\label{notation-fin-conj}
We write $\bigwedge \Gamma$, where $\Gamma$ is a finite set of formulas 
from $\SForm$ with enumeration 
${\langle A_i \rangle}_{i \in \set{1,\ldots,n}}$, 
for $A_1 \CAnd \ldots \CAnd A_n$.
\end{snot}
The formula that $\bigwedge \Gamma$ stands for is not uniquely 
determined.
Different choices of the enumeration lead to different formulas.
However, this does not matter because the formulas in question are the 
same up to associativity and commutativity of $\CAnd$.

\begin{snot}
\label{notation-syntac-eqv}
We write $e_1 \equiv e_2$, where $e_1$ and $e_2$ are terms from 
$\STerm$ or formulas from $\SForm$, to indicate that
$e_1$ is syntactically equal to $e_2$.
\end{snot}

\begin{snot}
\label{notation-FOCL}
We write \FOCL\ for the version of classical logic with the same 
language as $\foBDif$.
\end{snot}

\subsection{Free variables and substitution}
Free variables of a term or formula and substitution for variables in a 
term or formula are defined in the usual way.
\begin{snot}
\label{notation-substitution}
Let $x$ be a variable from $\SVar$, $t$ be a term from 
$\STerm$, and $e$ be a term from $\STerm$ or a formula 
from $\SForm$.
Then we write $\subst{x \assign t} e$ for the result of substituting the 
term $t$ for the free occurrences of the variable $x$ in~$e$, 
avoiding (by means of renaming of bound variables) free variables 
becoming bound in $t$.
\end{snot}

\subsection{Notational conventions}
The following will sometimes be used without mentioning (with or without 
decoration):
$x$~as a meta-variable ranging over all variables from 
$\SVar$,
$t$~as a meta-variable ranging over all terms from 
$\STerm$, 
$A$ as a meta-variable ranging over all formulas from 
$\SForm$, and
$\vGamma$ as a meta-variable ranging over all sets of formulas from
$\SForm$.

The string representation of terms and formulas suggested by the 
formation rules given above can lead to syntactic ambiguities. 
Parentheses are used to avoid  such ambiguities.
The need to use parentheses is reduced by ranking the precedence of the 
logical connectives $\Not$, $\CAnd$, $\COr$, $\IImpl$.
The enumeration presents this order from the highest precedence to the
lowest precedence.
Moreover, the scope of the quantifiers extends as far as possible to
the right and
$\CForall{x_1}{\cdots \CForall{x_n}{A}}$ and
$\CExists{x_1}{\cdots \CExists{x_n}{A}}$ are usually written 
as $\CForall{x_1,\ldots,x_n}{A}$ and $\CExists{x_1,\ldots,x_n}{A}$,
respectively.
\begin{snot}
\label{notation-abbreviations}
In what follows, the following abbreviations will be used: 
\[
\renewcommand{\arraystretch}{1.25}
\begin{array}[t]{r@{\;\;}c@{\;\;}l@{\;\;}l}
t_1 \neq t_2   & \mathrm{stands\; for} & \Not (t_1 = t_2), \\
\True          & \mathrm{stands\; for} & \Not \False.
\end{array}
\]
\end{snot}

\section{Logical Consequence in $\foBDif$}
\label{CONSEQUENCE}

The logical consequence relation of $\foBDif$ is defined using 
the logical matrix of $\foBDif$, the structures of 
$\foBDif$, the assignments in such a structure, and the 
valuation in such a structure under such an assigment.

\subsection{Matrix}
The interpretation of the logical symbols of $\foBDif$ is given by means 
of a logical matrix.

In the definition of this matrix, $\VTrue$ (\emph{true only}), 
$\VFalse$ (\emph{false only}), $\VBoth$ (\emph{both true and false}), 
and $\VNeither$ (\emph{neither true nor false}) are taken as truth 
values.
Moreover, use is made of the partial order $\leq$ on the set 
$\set{\VTrue,\VFalse,\VBoth,\VNeither}$ in which $\VFalse$ is the least 
element, $\VTrue$ is the greatest element, and $\VBoth$ and $\VNeither$ 
are incomparable.
We write $\inf V$ and $\sup V$, 
where $V \subseteq \set{\VTrue,\VFalse,\VBoth,\VNeither}$, for the
greatest lower bound and least upper bound, respectively, of $V$ with 
respect to~$\leq$.

\begin{sdef}
\label{def-matrix}
\sloppy
The \emph{matrix} of $\foBDif$ is the triple 
$\langle \TValue, \DValue, \TFunct \rangle$, where:%
\footnote
{We write $S^n$ for the $n$-fold cartesian power of the set $S$ and 
 $\pset(S)$ for the powerset of the set $S$.}
\begin{itemize}
\item
$\TValue = \set{\VTrue,\VFalse,\VBoth,\VNeither}$;
\item
$\DValue = \set{\VTrue,\VBoth}$;
\item
$\TFunct$ is the function with domain 
$\set{\False,\Not,\CAnd,\COr,\IImpl,\forall,\exists}$ such that
$\TFunct(\False) \mathbin{:} \TValue^0 \to \ \TValue$, 
$\TFunct(\Not) \mathbin{:} \TValue^1 \to \TValue$,\,\, 
$\TFunct(\CAnd), \TFunct(\COr), \TFunct(\IImpl) \mathbin{:}
 \TValue^2 \to \TValue$, and 
$\TFunct(\forall), \TFunct(\exists) \mathbin{:}
 \mathcal{P}(\TValue) \diff \set{\emptyset} \to \TValue$
and those functions are defined as follows:
\[
\begin{array}{rcl}
\TFunct(\False) & = &
 \begin{array}[t]{l}
 \VFalse \;,
 \end{array}
\\[.5ex]
\TFunct(\Not)(a) & = &
 \left \{
 \begin{array}{l@{\;\;}l}
 \VTrue  & \mathrm{if}\; a = \VFalse \\
 \VFalse & \mathrm{if}\; a = \VTrue \\
 a       & \mathrm{otherwise}\;,
 \end{array}
 \right.
\end{array}
\]
\[
\begin{array}{rcl}
\TFunct(\CAnd)(a_1,a_2)  & = &
 \begin{array}[t]{l}
 \inf \set{a_1,a_2}\;,
 \end{array}
\vspace*{.5ex} \\
\TFunct(\COr)(a_1,a_2)   & = &
 \begin{array}[t]{l}
 \sup \set{a_1,a_2}\;,
 \end{array}
\vspace*{.5ex} \\
\TFunct(\IImpl)(a_1,a_2) & = &
 \left \{
 \begin{array}{l@{\;\;}l}
 \VTrue  & \mathrm{if}\; a_1 \notin \set{\VTrue,\VBoth} \\
 a_2 & \mathrm{otherwise}\;,
 \end{array}
 \right.
\vspace*{.5ex} \\
\TFunct(\forall)(V) & = & \inf V\;,
\vspace*{.5ex} \\
\TFunct(\exists)(V) & = & \sup V\;,
\end{array}
\]
where 
$a$, $a_1$, and $a_2$ range over all truth values from $\TValue$ and 
$V$ ranges over all non-empty subsets of $\TValue$.
\end{itemize}
$\TValue$ is the set of \emph{truth values} of $\foBDif$,
$\DValue$ is the set of \emph{designated truth values} of 
$\foBDif$, and $\TFunct(\False)$, $\TFunct(\Not)$, 
$\TFunct(\CAnd)$, $\TFunct(\COr)$, $\TFunct(\IImpl)$, 
$\TFunct(\forall)$, and $\TFunct(\exists)$ are the 
\emph{truth functions} that are the interpretations of the logical 
symbols $\False$, $\Not$, $\CAnd$, $\COr$, $\IImpl$, $\forall$, and 
$\exists$, respectively.
The set of \emph{non-designated truth values} of $\foBDif$, 
written $\NDValue$, is the set~$\TValue \diff \DValue$.
\end{sdef}
The idea behind the designated truth values is that a formula is valid 
if its truth value with respect to all structures and assignments in 
those structures (both defined below) is a designated truth value. 

\subsection{Structures}

The possible interpretations of the non-logical symbols of $\foBDif$ are 
given by means of structures.
\begin{sdef}
\label{def-structure}
A \emph{structure} $\mathbf{A}$ of $\foBDif$ is a pair 
$\langle \mathcal{U}\sp\mathbf{A}, \mathcal{I}\sp\mathbf{A} \rangle$, 
where:
\begin{itemize}
\item
$\mathcal{U}\sp\mathbf{A}$ is a set, 
called the \emph{domain} of $\mathbf{A}$, 
such that $\mathcal{U}\sp\mathbf{A} \neq \emptyset$ and
$\mathcal{U}\sp\mathbf{A} \inter \TValue = \emptyset$;
\item
$\mathcal{I}\sp\mathbf{A}$ is a function with domain 
$\Union \set{\Func{n} \where n \in \Nat} \union
 \Union \set{\Pred{n} \where n \in \Nat}$
such that
\begin{itemize}
\item
$\mathcal{I}\sp\mathbf{A}(c) \in \mathcal{U}\sp\mathbf{A}$ 
for every $c \in \Func{0}$;
\item
$\mathcal{I}\sp\mathbf{A}(f) :
 {\mathcal{U}\sp\mathbf{A}}^{n+1} \to \mathcal{U}\sp\mathbf{A}$
for every $f \in \Func{n+1}$ and $n \in \Nat$; 
\item
$\mathcal{I}\sp\mathbf{A}(p) \in \TValue$ 
for every $p \in \Pred{0}$;
\item
$\mathcal{I}\sp\mathbf{A}(P) : 
 {\mathcal{U}\sp\mathbf{A}}^{n+1} \to \TValue$
for every $P \in \Pred{n+1}$ and $n \in \Nat$, \\
where, for all $d_1,d_2 \in \mathcal{U}\sp\mathbf{A}$,
$\mathcal{I}\sp\mathbf{A}(\Meq)(d_1,d_2) \in \DValue$ iff $d_1 = d_2$.
\end{itemize}
\end{itemize}
\end{sdef}

\subsection{Assignments}

The possible interpretations of the variables of $\foBDif$ are given by 
means of assignments.
\begin{sdef}
\label{def-assignment}
Let $\mathbf{A}$ be a structure of $\foBDif$.
Then an \emph{assignment} in $\mathbf{A}$ is a function
$\alpha: \SVar \to \mathcal{U}\sp\mathbf{A}$.
\end{sdef}
\begin{snot}
Let $\mathbf{A}$ be a structure of $\foBDif$, and
let $\alpha$ be an assignment in $\mathbf{A}$, $x \in \SVar$, 
and $d \in \mathcal{U}\sp\mathbf{A}$. 
Then we write $\alpha(x \to d)$ for the assignment $\alpha'$ in 
$\mathbf{A}$ such that $\alpha'(x) = d$ and $\alpha'(y) = \alpha(y)$ if 
$y \not\equiv x$.
\end{snot}

\subsection{Valuations}

The possible interpretations of the terms and formulas of \foBDif\ are 
given by means of valuations.
\begin{sdef}
\label{def-valuation}
Let $\mathbf{A}$ be a structure of $\foBDif$, and 
let $\alpha$ be an assignment in $\mathbf{A}$.
Then the \emph{valuation of terms in structure $\mathbf{A}$ 
under assignment $\alpha$} is a function 
$\Term{\ph}{\mathbf{A}}{\alpha} : \STerm \to \mathcal{U}\sp\mathbf{A}$ 
and the \emph{valuation of formulas in structure $\mathbf{A}$ 
under assignment $\alpha$} is a function 
$\Term{\ph}{\mathbf{A}}{\alpha} : \SForm \to \TValue$.
These valuation functions are inductively defined in 
Table~\ref{table-interpretation}.
\begin{table}[t!]
\caption{Valuations of terms and formulas of $\foBDif$}
\label{table-interpretation}
\centering
$
\renewcommand{\arraystretch}{1.25}
\begin{array}[c]{@{}rcl@{}}
\hline
\\[-3ex]
\Term{x}{\mathbf{A}}{\alpha} & = &
 \begin{array}[t]{l}
 \alpha(x) \;,
 \end{array}
\\[1.5ex]
\Term{c}{\mathbf{A}}{\alpha} & = &
 \begin{array}[t]{l}
 \mathcal{I}\sp\mathbf{A}(c) \;,
 \end{array}
\\[.5ex]
\Term{f(t\sb1,\ldots,t\sb{n+1})}{\mathbf{A}}{\alpha} & = &
 \begin{array}[t]{l}
 \mathcal{I}\sp\mathbf{A}(f)(\Term{t\sb1}{\mathbf{A}}{\alpha},\ldots,
                             \Term{t\sb{n+1}}{\mathbf{A}}{\alpha})
 \end{array}
\\[1.5ex]
\Term{p}{\mathbf{A}}{\alpha} & = &
 \begin{array}[t]{l}
 \mathcal{I}\sp\mathbf{A}(p) \;,
 \end{array}
\\[.5ex]
\Term{P(t\sb1,\ldots,t\sb{n+1})}{\mathbf{A}}{\alpha} & = &
 \begin{array}[t]{l}
 \mathcal{I}\sp\mathbf{A}(P)(\Term{t\sb1}{\mathbf{A}}{\alpha},\ldots,
                \Term{t\sb{n+1}}{\mathbf{A}}{\alpha}) \;,
 \end{array}
\\[1.5ex]
\Term{\False}{\mathbf{A}}{\alpha} & = & \TFunct(\False) \;,
\\[.5ex]
\Term{\Not A}{\mathbf{A}}{\alpha} & = & 
\TFunct(\Not)(\Term{A}{\mathbf{A}}{\alpha}) \;,
\\[.5ex]
\Term{A\sb1 \CAnd A\sb2}{\mathbf{A}}{\alpha} & = &
\TFunct(\CAnd)(\Term{A\sb1}{\mathbf{A}}{\alpha},
                  \Term{A\sb2}{\mathbf{A}}{\alpha}) \;,
\\[.5ex]
\Term{A\sb1 \COr A\sb2}{\mathbf{A}}{\alpha} & = &
\TFunct(\COr)(\Term{A\sb1}{\mathbf{A}}{\alpha},
                 \Term{A\sb2}{\mathbf{A}}{\alpha}) \;,
\\[.5ex]
\Term{A\sb1 \IImpl A\sb2}{\mathbf{A}}{\alpha} & = &
\TFunct(\IImpl)(\Term{A\sb1}{\mathbf{A}}{\alpha},
                   \Term{A\sb2}{\mathbf{A}}{\alpha}) \;,
\\[.5ex]
\Term{\CForall{x}{A}}{\mathbf{A}}{\alpha} & = &
\TFunct(\forall)(\set{\Term{A}{\mathbf{A}}{\alpha(x \to d)} \where
                         d \in \mathcal{U}\sp\mathbf{A}}) \;,
\\[.5ex]
\Term{\CExists{x}{A}}{\mathbf{A}}{\alpha} & = &
\TFunct(\exists)(\set{\Term{A}{\mathbf{A}}{\alpha(x \to d)} \where
                         d \in \mathcal{U}\sp\mathbf{A}}) \;,
\\[.5ex]
\hline
\end{array}
$
\end{table}
In this table, 
$x$ ranges over all variables from $\SVar$, 
$c$~ranges over all function symbols from $\Func{0}$,
$f$ ranges over all function symbols from $\Func{n+1}$, 
$p$ ranges over all predicate symbols from $\Pred{0}$,
$P$ ranges over all  predicate symbols from $\Pred{n+1}$, 
$t_1$, \ldots, $t_{n+1}$ range over all terms from $\STerm$, 
and 
$A$, $A_1$, and $A_2$ range over all formulas from~$\SForm$.
\end{sdef}

\subsection{Logical consequence}

The logical consequence relation of \foBDif\ is defined in terms of
valuations of formulas.
\begin{sdef}
\label{def-LCon}
\sloppy
Let $\vGamma$ and $\vDelta$ be sets of formulas from $\SForm$.
Then 
\emph{$\vDelta$ is a logical \linebreak[2] consequence} of $\vGamma$, 
written $\vGamma \LCon \vDelta$, iff
for all structures $\mathbf{A}$ of $\foBDif$,
for all assignments $\alpha$ in $\mathbf{A}$,
if $\Term{A}{\mathbf{A}}{\alpha} \in \DValue$ for all $A \in \vGamma$, 
then $\Term{A'}{\mathbf{A}}{\alpha} \in \DValue$ for some 
$A' \in \vDelta$.
\end{sdef}
\begin{snot}
\label{notation-LCon}
We write ${} \LCon \Delta$, where $\Delta \subseteq \SForm$, for 
$\emptyset \LCon \Delta$. 
\end{snot}

\begin{snot}
\label{notation-notLCon}
We write $\vGamma \notLCon \vDelta$, 
where $\Gamma,\Delta \subseteq \SForm$, for not $\vGamma \LCon \vDelta$.
\end{snot}

\begin{snot}
\label{notation-form-set}
We write $\Gamma,\Gamma'$, where $\Gamma,\Gamma' \subseteq \SForm$, for 
$\Gamma \union \Gamma'$ and $A$, where $A \in \SForm$, for $\set{A}$ 
wherever the context expects a subset of $\SForm$.
\end{snot}

The following two propositions follow easily from the definition of the 
logical consequence relation of $\foBDif$.
The first proposition concerns properties of the logical consequence 
relation of $\foBDif$ that emphasize the similarities with 
the logical consequence relation of \FOCL.
The second proposition concerns properties of the logical consequence 
relation of $\foBDif$ that emphasize the dissimilarities with 
the logical consequence relation of \FOCL.
\begin{proposition}
\label{prop-normal-contained}
\mbox{} 
\begin{enumerate}
\item
\foBDif\ is \emph{normal}, i.e.\ $\LCon$ is such that for all 
$\vGamma, \vDelta \subseteq \SForm$, $A_1, A_2 \in \SForm$, 
and $x,y \in \SVar$:
\[
\renewcommand{\arraystretch}{1.25}
\begin{array}[t]{r@{\;}c@{\;}l}
\vGamma \LCon \vDelta, A_1 \CAnd A_2    & \mathrm{iff} &
\vGamma \LCon \vDelta, A_1 \;\mathrm{and}\;
\vGamma \LCon \vDelta, A_2\;,
\\
A_1 \COr A_2, \vGamma \LCon \vDelta     & \mathrm{iff} &
A_1, \vGamma \LCon \vDelta \;\mathrm{and}\;
A_2, \vGamma \LCon \vDelta\;,
\\
\vGamma \LCon \vDelta, A_1 \IImpl A_2   & \mathrm{iff} & 
A_1, \vGamma \LCon \vDelta, A_2\;,
\\
\vGamma \LCon \vDelta, \CForall{x}{A_1} & \mathrm{iff} &
\vGamma \LCon \vDelta, \subst{x \assign y}A_1 
\\
& & \qquad
 \mbox{provided $y$ is not free in $\vGamma \union \vDelta \union \set{A_1}$}\;,
\\
\CExists{x}{A_1}, \vGamma \LCon \vDelta & \mathrm{iff} &
\subst{x \assign y}A_1, \vGamma \LCon \vDelta 
\\
& & \qquad
 \mbox{provided $y$ is not free in $\vGamma \union \vDelta \union \set{A_1}$}\;,
\end{array}
\]
\item
\foBDif\ is $\Not$-\emph{contained in classical logic}, i.e.\
there exists a logic with the same language as $\foBDif$ and 
a logical consequence relation $\LCon'$ such that:
\begin{itemize}
\item
${\LCon} \subseteq {\LCon'}$;
\item
$\LCon'$ is induced by a matrix 
$\langle \TValue', \DValue', \TFunct' \rangle$  such that 
$\TValue' = \set{\VTrue,\VFalse}$, 
$\DValue' = \set{\VTrue}$, and $\TFunct'(\Not)$ is defined as follows:
\[
\renewcommand{\arraystretch}{1.25}
\begin{array}[t]{c}
\TFunct'(\Not)(a) =
 \left \{
 \begin{array}{l@{\;\;}l}
 \VTrue  & \mathrm{if}\; a = \VFalse \\
 \VFalse & \mathrm{if}\; a = \VTrue\;,
 \end{array}
 \right.
\end{array}
\]
where $a$ ranges over all truth values in $\TValue'$.
\end{itemize}
\end{enumerate}
\end{proposition}
\pagebreak[2]
\begin{proposition}
\label{prop-paradefinite}
\mbox{} 
\begin{enumerate}
\item
there exist a $\Gamma \subseteq \SForm$ and $A, A' \in \SForm$ such that 
$\Gamma \LCon A$ and $\Gamma \LCon \Not A$, but $\Gamma \not\LCon A'$;
\item
there exist a $\vGamma \subseteq \SForm$ and
$A, A' \in \SForm$ such that $\vGamma, A \LCon A'$ and 
$\vGamma, \Not A \LCon A'$, but $\vGamma \notLCon A'$.
\end{enumerate}
\end{proposition}
Because $\foBDif$ is normal and $\Not$-contained in classical logic, 
property~1 and~2 of Proposition~\ref{prop-paradefinite} imply that 
$\foBDif$ is \emph{paraconsistent} and \emph{paracomplete}, 
respectively, in the sense of~\cite{AA17a}.

$\foBDif$ is not maximally paraconsistent in the sense of~\cite{AA17a}
because a logic obtained by extending the logical consequence relation 
of $\foBDif$ is not paraconsistent or not paracomplete, but not 
necessarily both. 
The following proposition, which follows easily from the definition of 
the logical consequence relation of $\foBDif$, concerns another property 
about the extent to which $\foBDif$ is paraconsistent.
\begin{proposition}
\label{prop-fully-paraconsistent}
Let $\SForm'$ be the set of all formulas from $\SForm$ in which 
$\False$ does not occur.
Then, for all $\Gamma \subset \SForm'$, there exists an $A \in \SForm'$ 
such that $\Gamma \notLCon A$.
\end{proposition}

\section{A Proof System for $\foBDif$}
\label{PROOF-SYSTEM}

A proof system for \foBDif\ is presented that is sound and complete 
with respect to the logical consequence relation defined in
Section~\ref{CONSEQUENCE}.

\subsection{Sequents and rules of inference}
\label{subsect-proof-system}

The presented proof system for $\foBDif$ is a sequent calculus proof 
system.
First, we define what a sequent of $\foBDif$ is.

\begin{sdef}
\label{def-sequent}
A \emph{sequent} of $\foBDif$ is an expression of the form 
$\Gamma \scEnt \Delta$, where $\Gamma$ and $\Delta$ are finite sets of 
formulas from $\SForm$.
\end{sdef}
\begin{snot}
\label{notation-sequent}
We write ${} \scEnt \Delta$, where $\Delta \subseteq \SForm$, for 
$\emptyset \scEnt \Delta$. 
\end{snot}
In the sequel, it will be proved that a sequent $\Gamma \scEnt \Delta$ 
can be proved by means of the rules of inference of the sequent calculus 
proof system for $\foBDif$ iff $\Gamma \LCon \Delta$ holds.
\begin{sdef}
\label{def-proof-system}
The \emph{sequent calculus proof system} for $\foBDif$ consists of the 
inference rules given in Table~\ref{table-proof-system}.
\begin{table}[!p]
\caption{Inference rules of a sequent calculus proof system for $\foBDif$}
\label{table-proof-system}
\vspace*{-2ex} 
\renewcommand{\arraystretch}{1.2} 
\centering
\begin{tabular}[t]{@{}c@{}}
\hline
\\[-2.5ex]
\begin{small}
\begin{tabular}{@{}l@{}}
\InfRule{Id}
 {{}}
 {A, \Gamma \scEnt \Delta, A}
\\[3ex] 
\InfRule{$\False$-L}
 {{}}
 {\False, \Gamma \scEnt \Delta}
\\[3ex] 
\InfRule{$\CAnd$-L}
 {A\sb1, A\sb2, \Gamma \scEnt \Delta}
 {A\sb1 \CAnd A\sb2, \Gamma \scEnt \Delta}
\\[3ex]
\InfRule{$\COr$-L}
 {A\sb1, \Gamma \scEnt \Delta \quad
  A\sb2, \Gamma \scEnt \Delta}
 {A\sb1 \COr A\sb2, \Gamma \scEnt \Delta}
\\[3ex]
\InfRule{$\IImpl$-L}
 {\Gamma \scEnt \Delta, A\sb1 \quad
  A\sb2, \Gamma \scEnt \Delta}
 {A\sb1 \IImpl A\sb2, \Gamma \scEnt \Delta}
\\[3ex]
\InfRule{$\forall$-L}
 {\subst{x \assign t}A, \Gamma \scEnt \Delta}
 {\CForall{x}{A}, \Gamma \scEnt \Delta}
\\[3ex]
\InfRuleC{$\exists$-L}
 {\subst{x \assign y}A, \Gamma \scEnt \Delta}
 {\CExists{x}{A}, \Gamma \scEnt \Delta}
 {$\ast$}
\\[3ex]
\InfRule{$\Not \Not$-L}
 {A, \Gamma \scEnt \Delta}
 {\Not \Not A, \Gamma \scEnt \Delta}
\\[3ex]
\InfRule{$\Not \CAnd$-L}
 {\Not A\sb1, \Gamma \scEnt \Delta \quad
  \Not A\sb2, \Gamma \scEnt \Delta}
 {\Not (A\sb1 \CAnd A\sb2), \Gamma \scEnt \Delta}
\\[3ex]
\InfRule{$\Not \COr$-L}
 {\Not A\sb1, \Not A\sb2, \Gamma \scEnt \Delta}
 {\Not (A\sb1 \COr A\sb2), \Gamma \scEnt \Delta}
\\[3ex]
\InfRule{$\Not \IImpl$-L}
 {A\sb1, \Not A\sb2, \Gamma \scEnt \Delta}
 {\Not (A\sb1 \IImpl A\sb2), \Gamma \scEnt \Delta}
\\[3ex]
\InfRuleC{$\Not \forall$-L}
 {\Not \subst{x \assign y}A, \Gamma \scEnt \Delta}
 {\Not \CForall{x}{A}, \Gamma \scEnt \Delta}
 {$\ast$}
\\[3ex]
\InfRule{$\Not \exists$-L}
 {\Not \subst{x \assign t}A, \Gamma \scEnt \Delta}
 {\Not \CExists{x}{A}, \Gamma \scEnt \Delta}
\\[3ex]
\InfRule{$=$-Refl}
 {t = t, \Gamma \scEnt \Delta}
 {\Gamma \scEnt \Delta}
\\[3ex]
\end{tabular}
\qquad \qquad
\begin{tabular}{@{}l@{}} 
\InfRule{Cut}
 {\Gamma \scEnt \Delta, A \quad
  A, \Gamma' \scEnt \Delta'}
 {\Gamma', \Gamma \scEnt \Delta, \Delta'}
\\[3ex] 
\InfRule{$\Not \False$-R}
 {{}}
 {\Gamma \scEnt \Delta, \Not \False}
\\[3ex] 
\InfRule{$\CAnd$-R}
 {\Gamma \scEnt \Delta, A\sb1 \quad
  \Gamma \scEnt \Delta, A\sb2}
 {\Gamma \scEnt \Delta, A\sb1 \CAnd A\sb2}
\\[3ex] 
\InfRule{$\COr$-R}
 {\Gamma \scEnt \Delta, A\sb1, A\sb2}
 {\Gamma \scEnt \Delta, A\sb1 \COr A\sb2}
\\[3ex]
\InfRule{$\IImpl$-R}
 {A\sb1, \Gamma \scEnt \Delta, A\sb2}
 {\Gamma \scEnt \Delta, A\sb1 \IImpl A\sb2}
\\[3ex]
\InfRuleC{$\forall$-R}
 {\Gamma \scEnt \Delta, \subst{x \assign y}A}
 {\Gamma \scEnt \Delta, \CForall{x}{A}}
 {$\ast$}
\\[3ex]
\InfRule{$\exists$-R}
 {\Gamma \scEnt \Delta, \subst{x \assign t}A}
 {\Gamma \scEnt \Delta, \CExists{x}{A}}
\\[3ex]
\InfRule{$\Not \Not$-R}
 {\Gamma \scEnt \Delta, A}
 {\Gamma \scEnt \Delta, \Not \Not A}
\\[3ex] 
\InfRule{$\Not \CAnd$-R}
 {\Gamma \scEnt \Delta, \Not A\sb1, \Not A\sb2}
 {\Gamma \scEnt \Delta, \Not (A\sb1 \CAnd A\sb2)}
\\[3ex] 
\InfRule{$\Not \COr$-R}
 {\Gamma \scEnt \Delta, \Not A\sb1 \quad
  \Gamma \scEnt \Delta, \Not A\sb2}
 {\Gamma \scEnt \Delta, \Not (A\sb1 \COr A\sb2)}
\\[3ex] 
\InfRule{$\Not \IImpl$-R}
 {\Gamma \scEnt \Delta, A\sb1 \quad
  \Gamma \scEnt \Delta, \Not A\sb2}
 {\Gamma \scEnt \Delta, \Not (A\sb1 \IImpl A\sb2)}
\\[3ex]
\InfRule{$\Not \forall$-R}
 {\Gamma \scEnt \Delta, \Not \subst{x \assign t}A}
 {\Gamma \scEnt \Delta, \Not \CForall{x}{A}}
\\[3ex]
\InfRuleC{$\Not \exists$-R}
 {\Gamma \scEnt \Delta, \Not \subst{x \assign y}A}
 {\Gamma \scEnt \Delta, \Not \CExists{x}{A}}
 {$\ast$}
\\[3ex]
\InfRule{$=$-Repl}
 {\subst{x \assign t\sb1}A, \Gamma \scEnt \Delta}
 {t\sb1 = t\sb2, \subst{x \assign t\sb2}A, \Gamma \scEnt \Delta}
\\[3ex]
\end{tabular}
\end{small}
\\[3ex]
\begin{tabular}{@{}l@{}}
$\ast$ 
provided $y$ is not free in $\Gamma \union \Delta \union \set{A}$.
\vspace*{1ex} \par
\end{tabular}
\\
\hline
\end{tabular}
\end{table}
In this table, 
$x$ and $y$ range over all variables from $\SVar$,\,
$t$, $t_1$, and $t_2$ range over all terms from $\STerm$,\, 
$A$, $A_1$, and $A_2$ range over all formulas from $\SForm$, and
$\Gamma$ and $\Delta$ range over all finite sets of formulas from 
$\SForm$. 
\end{sdef}
In the sequel, inference rules with zero premises will also be called 
axioms.

\subsection{Proofs of sequents}

Below, we make precise what counts as proof of a sequent of $\foBDif$ by 
means of the inference rules of the sequent calculus proof system for 
$\foBDif$.
\begin{sdef}
\label{def-proof}
In the sequent calculus proof system for $\foBDif$, 
a \emph{proof of a sequent $\Gamma \scEnt \Delta$} is a finite sequence 
$\seq{s_1,\ldots,s_n}$ of sequents such that $s_n$ equals 
$\Gamma \scEnt \Delta$ and, for each 
$i \in \set{1,\ldots,n}$, the following condition holds:
\begin{itemize}
\item
$s_i$ is the conclusion of an instance of some inference rule from the 
proof system of $\foBDif$ whose premises are among 
$s_1,\ldots,s_{i-1}$.
\end{itemize}
A sequent $\Gamma \scEnt \Delta$ is said to be \emph{provable} iff there 
exists a proof of $\Gamma \scEnt \Delta$.
\end{sdef}

The following definition is useful in formulating the soundness and 
completeness result for the sequent calculus proof system for $\foBDif$.
\begin{sdef}
\label{def-derivation}
Let $\Gamma,\Delta \subseteq \SForm$.
Then $\Delta$ is \emph{derivable} from $\Gamma$, written 
$\Gamma \LDer \Delta$, iff there exist finite sets 
$\Gamma' \subseteq \Gamma$ and $\Delta' \subseteq \Delta$ such that the 
sequent $\Gamma' \scEnt \Delta'$ is provable.
\end{sdef}

The sequent calculus proof system for $\foBDif$ is sound and complete 
with respect to the logical consequence relation $\LCon$ of $\foBDif$.
\begin{theorem}
\label{theorem-sound-complete-BD}
For all $\Gamma,\Delta \subseteq \SForm$,
$\Gamma \LDer \Delta \;\; \mathrm{iff} \;\; \Gamma \LCon \Delta$.
\end{theorem}
\begin{proof}
See Appendix~\ref{SOUND-COMPLETE}.%
\footnote
{In Appendix~\ref{SOUND-COMPLETE}, Theorem~\ref{theorem-sound-complete-BD} 
 is proved together with a cut-elimination result for the sequent 
 calculus proof system of $\foBDif$.}
\qed
\end{proof}

\subsection{Extensions of the proof system}
\label{subsect-FOCL}

The languages of $\FOCL$ and $\foBDif$ are the same.
A sound and complete sequent calculus proof system for $\FOCL$ 
can be obtained by adding two inference rules to the presented
sequent calculus proof system of $\foBDif$.

\begin{snot}
We write $\clLCon$ for the logical consequence relation of $\FOCL$.%
\footnote
{In this paper, the logical consequence relation of \FOCL\ is assumed to 
 be known.}
\end{snot}
\begin{theorem}
\label{theorem-sound-complete-CL}
A sequent calculus proof system for $\FOCL$ that is sound and complete
with respect to $\clLCon$ can be obtained by adding the following 
inference rules to the inference rules given in 
Table~\ref{table-proof-system}:%
\\[1.5ex]
\mbox{} \hfill
\begin{tabular}{@{}c@{}} 
\InfRule{$\Not$-L}
 {\Gamma \scEnt \Delta, A}
 {\Not A, \Gamma \scEnt \Delta}\;,
\hspace*{4em}
\InfRule{$\Not$-R}
 {A, \Gamma \scEnt \Delta}
 {\Gamma \scEnt \Delta, \Not A}\;.
\end{tabular}
\hfill \vspace*{1.5ex}
\end{theorem}
\begin{proof}
By the addition of the above two inference rules, the inference rules 
given in Table~\ref{table-proof-system} whose name begins with $\Not$ 
become derived inference rules.
Moreover, $\scEnt t = t$ and
$t_1 = t_2, \subst{x \assign t_1}A \scEnt \subst{x \assign t_2}A$ are 
derived axioms of the presented proof system of $\FOCL$ and the 
inference rules $=$-Refl and $=$-Repl are derived inference rules of 
the proof system obtained by replacing them by those axioms.
So, we can remove the inference rules from Table~\ref{table-proof-system} 
whose name begins with $\Not$ and replace the inference rules $=$-Refl 
and $=$-Repl by the above axioms.
The sequent calculus proof system so obtained is a well-known sound and 
complete sequent calculus proof system for $\FOCL$.
\qed
\end{proof}
If we add only the inference rule $\Not$-R to the sequent calculus proof 
system of $\foBDif$, then we obtain a sound and complete proof 
system of the paraconsistent (but not paracomplete) logic \foLPif\ 
presented in~\cite{Mid22b}.
If we add only the inference rule $\Not$-L to the sequent calculus proof 
system of $\foBDif$, then we obtain a sound and complete proof 
system of the obvious first-order version of the paracomplete (but not 
paraconsistent) logic $\Klif$ presented in~\cite{Mid17a}.

\section{Logical Equivalence in \foBDif}
\label{EQUIVALENCE}

The laws of logical equivalence that the logical equivalence relation of 
$\foBDif$ satisfies constitute a potentially relevant property of 
$\foBDif$.  
\begin{sdef}
\label{def-LEqv}
Let $A_1$ and $A_2$ be formulas from $\SForm$. 
Then $A_1$ is \emph{logically equivalent} to $A_2$, written 
$A_1 \LEqv A_2$, iff,
for all structures $\mathbf{A}$ of $\foBDif$,
for all assignments $\alpha$ in $\mathbf{A}$,
$\Term{A_1}{\mathbf{A}}{\alpha} = \Term{A_2}{\mathbf{A}}{\alpha}$.
\end{sdef}

The following theorem concerns classical laws of logical equivalence
that are satisfied by the logical equivalence relation of \foBDif.
\begin{theorem}
\label{theorem-soundness}
The logical equivalence relation of \foBDif\ satisfies laws (1)--(15) 
from Table~\ref{laws-lequiv}.
\begin{table}[!t]
\caption{The distinguishing laws of logical equivalence for \foBDif}
\label{laws-lequiv}
\begin{eqntbl}
\begin{neqncol}
(1)  & A \CAnd \False \LEqv \False \\
(3)  & A \CAnd \True \LEqv A \\
(5)  & A \CAnd A \LEqv A \\
(7)  & A_1 \CAnd A_2 \LEqv A_2 \CAnd A_1 \\
(9)  & \Not (A_1 \CAnd A_2) \LEqv \Not A_1 \COr \Not A_2 \\
(11) & \Not \Not A \LEqv A \\
(12) & (A_1 \CAnd (A_1 \IImpl \False)) \IImpl A_2 \LEqv \True \\
(14) & \CForall{x}{(A_1 \CAnd A_2)} \LEqv (\CForall{x}{A_1}) \CAnd A_2 \\ 
     & \hfill \mbox{if $x$ is not free in $A_2$} 
\end{neqncol}
\qquad
\begin{neqncol}
(2)  & A \COr \True \LEqv \True \\
(4)  & A \COr \False \LEqv A \\
(6)  & A \COr A \LEqv A \\
(8)  & A_1 \COr A_2 \LEqv A_2 \COr A_1 \\
(10) & \Not (A_1 \COr A_2) \LEqv \Not A_1 \CAnd \Not A_2 \\
     & \\
(13) & (A_1 \COr (A_1 \IImpl \False)) \IImpl A_2 \LEqv A_2 \\
(15) & \CExists{x}{(A_1 \COr A_2)} \LEqv (\CExists{x}{A_1}) \COr A_2 \\ 
     & \hfill \mbox{if $x$ is not free in $A_2$}  
\end{neqncol}
\end{eqntbl}
\end{table}
\end{theorem}
\begin{proof}
The proof is easy by constructing, for each of the laws concerned, truth 
tables for both sides.
\qed
\end{proof}

Laws (1)--(11) from Table~\ref{laws-lequiv} are the identity, 
annihilation, idempotent, commutative, and De~Morgan's laws for 
conjunction and disjunction and the double negation law known from 
classical logic. 
In the case of \FOCL, laws~(12) and~(13) from Table~\ref{laws-lequiv} 
\pagebreak[2]
follow from the following classical law of logical equivalence:
$\Not (A_1 \IImpl A_2) \LEqv A_1 \CAnd \Not A_2$.
However, this laws does not hold in the case of \foBDif.
That laws~(12) and~(13) hold in the case of \foBDif\ is easy to see 
given that
\[
\TFunct(\CAnd)(a, \TFunct(\IImpl)(a,\VFalse)) =
 \left \{
 \begin{array}{l@{\;\;}l}
 \VFalse   & \mathrm{if}\; a \neq \VNeither \\
 \VNeither & \mathrm{otherwise}\;
 \end{array}
 \right.
\;\; 
\TFunct(\COr)(a, \TFunct(\IImpl)(a,\VFalse)) =
 \left \{
 \begin{array}{l@{\;\;}l}
 \VTrue    & \mathrm{if}\; a \neq \VBoth \\
 \VBoth    & \mathrm{otherwise}\;.
 \end{array}
 \right.
\]

The following proposition follows easily from the definition of the 
matrix of $\foBDif$ and is used in the proof of the next theorem.
\begin{proposition}
\label{prop-matrix}
The matrix of $\foBDif$ has the following properties:
\begin{itemize}
\item
it is \emph{four-valued}: \\
$\TValue = \set{\VTrue,\VFalse,\VBoth,\VNeither}$ and  
$\DValue = \set{\VTrue,\VBoth}$;
\item
it is \emph{regular}: \\
the domain of $\TFunct$ is 
$\set{\False,\Not,\CAnd,\COr,\IImpl,\forall,\exists}$\;, \\
$
\renewcommand{\arraystretch}{1.25}
\begin{array}[t]{@{}l@{\;}c@{\;}l@{}} 
\TFunct(\False) = \VFalse\;,           
\\
\TFunct(\Not)(a) \in \DValue         & \mathrm{iff} & 
a \in \set{\VFalse,\VBoth}\;,
\\
\TFunct(\CAnd)(a_1,a_2) \in \DValue  & \mathrm{iff} &
a_1 \in \DValue \;\mathrm{and}\; a_2 \in \DValue\;, 
\\
\TFunct(\COr)(a_1,a_2) \in \DValue   & \mathrm{iff} &
a_1 \in \DValue \;\mathrm{or}\; a_2 \in \DValue\;, 
\\
\TFunct(\IImpl)(a_1,a_2) \in \DValue & \mathrm{iff} &
a_1 \in \NDValue \;\mathrm{or}\; a_2 \in \DValue\;,
\\
\TFunct(\forall)(V) \in \DValue & \mathrm{iff} &
V \inter \NDValue = \emptyset\;,
\\
\TFunct(\exists)(V) \in \DValue & \mathrm{iff} &
V \inter \DValue \neq \emptyset\;; 
\end{array}
$
\item 
it is \emph{classically closed}: \\
$\TFunct(\Not)(a_1)$,
$\TFunct(\CAnd)(a_1,a_2)$,
$\TFunct(\COr)(a_1,a_2)$, 
$\TFunct(\IImpl)(a_1,a_2) \in \set{\VTrue,\VFalse}$ if 
$a_1,a_2 \in \set{\VTrue,\VFalse}$,
$\TFunct(\forall)(V)$,
$\TFunct(\exists)(V) \in \set{\VTrue,\VFalse}$ if 
$V \subseteq \set{\VTrue,\VFalse}$.
\end{itemize}
\end{proposition}

Now suppose that in the matrix of \foBDif\ one or more of the functions
\mbox{$\TFunct(\False) \mathbin{:} \TValue^0 \to \TValue$},\,\, 
$\TFunct(\Not) \mathbin{:} \TValue^1 \to \TValue$,\,\, 
$\TFunct(\CAnd), \TFunct(\COr), \TFunct(\IImpl) \mathbin{:}
 \TValue^2 \to \TValue$, and 
$\TFunct(\forall), \TFunct(\exists) \mathbin{:}
 \mathcal{P}(\TValue) \diff \set{\emptyset} \to \TValue$
are changed.
Then those changes give rise to a different logic with a different 
logical equivalence relation.
Among the logics whose matrix is four-valued, regular, and classically 
closed, \foBDif\ is the only one whose logical equivalence relation 
satisfies all laws given in Table~\ref{laws-lequiv}.
\begin{theorem}
\label{theorem-uniqueness}
There is exactly one logic whose matrix is four-valued, regular, and 
classically closed and whose logical equivalence relation satisfies 
laws \mbox{(1)--(15)} from Table~\ref{laws-lequiv}.
\end{theorem}
\begin{proof}
See Appendix~\ref{UNIQUENESS}.
\qed
\end{proof}

It follows immediately from the proof of 
Theorem~\ref{theorem-uniqueness} that all proper subsets of laws 
(1)--(15) from Table~\ref{laws-lequiv} are insufficient to distinguish 
\foBDif\ completely from the other logics whose matrix is four-valued, 
regular, and classically closed.

\section{Naturalness of \foBDif\ Relative to \FOCL}
\label{NATURAL}

It is argued that \foBDif\ is the most natural paradefinite logic 
relative to \FOCL.

\FOCL\ and \foBDif\ have the same connectives, namely
$\False$, $\Not$, $\CAnd$, $\COr$, and $\IImpl$.
Arguments for the choice of connectives are:
\begin{itemize}
\item
any expansion of BD must include the connectives $\CAnd$, $\COr$, and 
$\Not$ because these are the connectives of BD;
\item
the expansion of BD with both the connectives $\IImpl$ and $\False$ has 
greater expressive power than an expansion of BD with only one of them;
\item
an expansion of BD with the connectives $\IImpl$ and $\False$ and other 
connectives known from classical logic does not have more expressive 
power than the expansion of BD with only the connectives $\IImpl$ and 
$\False$;
\item
an expansion of BD with connectives not known from classical logic 
does not deserve to be qualified as the most natural paradefinite logic 
relative to classical logic. 
\end{itemize}
Moreover, this choice of connectives yields a suitable language for 
the most natural paradefinite logic relative to classical logic: 
it guarantees that, for each connective available or definable in the 
version of classical logic other than $\Not$, a connective with the same 
properties with respect to logical consequence is available or definable 
in the paradefinite logic.

It is worth mentioning here that, although the connectives of BD are 
$\CAnd$, $\COr$, and~$\Not$, the falsity connective $\False$ and the 
implication connective $\IImpl$ for which the standard inference theorem 
holds are not definable in BD.
A relatively unknown consequence of expanding BD with the connectives 
$\IImpl$ and $\False$ is that several interesting connectives not known 
from classical logic become definable (see~\cite{Mid24b}, Section~6).
Additional connectives not known from classical logic are needed to 
obtain an expansion of BD with more expressive power than the 
expansion of BD with the connectives $\IImpl$ and $\False$.

It holds that 
(a)~for all $A \in \SForm$, $\Not A, A \clLCon \False$ and 
(b)~for all $A \in \SForm$, $\Not \False \clLCon A, \Not A$.
Property~(a) represent the \emph{law of non-contradiction} (LNC) and 
property~(b) represents the \emph{law of excluded middle} (LEM).
LNC is the only reason why \FOCL\ cannot serve as the underlying logic 
for theories that are inconsistent and LEM is the only reason why \FOCL\ 
cannot serve as the underlying logic for theories that are incomplete.
Getting rid of LNC and LEM is all that is needed to obtain a logic that 
can serve as the underlying logic for theories that are inconsistent or 
incomplete.

\foBDif\ can be thought as obtained in exactly this way.
That is, ${\clLCon} \diff {\LCon}$ is precisely the set of all classical 
logical consequences that exist due to either LNC or LEM.
This is made clear by the following proposition, which follows easily 
from the following (cf.~\cite{Mid26a}):
\begin{itemize}
\item
the sequent calculus proof systems of \foBDif\ and \FOCL\ presented in 
Section~\ref{PROOF-SYSTEM} are sound and complete with respect to the
logical consequence relations \linebreak[2] $\LCon$ and $\clLCon$, 
respectively (Theorems~\ref{theorem-sound-complete-BD} 
and~\ref{theorem-sound-complete-CL});
\item
the common inference rules of both proof systems that concern the 
connectives and quantifiers are all invertible;
\item
the common inference rules of both proof systems that concern the 
equality predicate are replaceable by the axioms $\scEnt t_1 = t_1$ and 
$t_1 = t_2,
 \subst{x \assign t_1}A_1 \scEnt \linebreak[2] \subst{x \assign t_2}A_1$
(see Theorem~\ref{theorem-sound-complete-CL});
\item
the inference rules $\Not$-L and $\Not$-R of the proof system of \FOCL\ 
are replaceable by the axioms $\Not A_1, A_1 \scEnt \False$ and 
$\Not \False \scEnt A_1, \Not A_1$.
\end{itemize}
\begin{proposition}
\label{proposition-LCon-BD-CL}
The logical consequence relation $\LCon$ of \foBDif\ is the smallest 
logical consequence relation that satisfies the following conditions 
for all $\Gamma, \Delta \subseteq \SForm$, $A_1, A_2 \in \SForm$, 
$t_1, t_2 \in \STerm$, and $x \in \SVar$:
\[
\renewcommand{\arraystretch}{1.25}
\begin{array}[t]{r@{\;\;}c@{\;\;}l}
\False, \Gamma \LCon \Delta & \!\!\!\!,
\\
\Gamma \LCon \Delta, \Not A_1         & \mathrm{iff} &
A_1, \Gamma \LCon \Delta\;,
\\
\Gamma \LCon \Delta, A_1 \CAnd A_2    & \mathrm{iff} &
\Gamma \LCon \Delta, A_1 \;\mathrm{and}\;
\Gamma \LCon \Delta, A_2\;,
\\
A_1 \COr A_2, \Gamma \LCon \Delta     & \mathrm{iff} &
A_1, \Gamma \LCon \Delta \;\mathrm{and}\;
A_2, \Gamma \LCon \Delta\;,
\\
\Gamma \LCon \Delta, A_1 \IImpl A_2   & \mathrm{iff} & 
A_1, \Gamma \LCon \Delta, A_2\;,
\\
\vGamma \LCon \vDelta, \CForall{x}{A_1} & \mathrm{iff} &
\vGamma \LCon \vDelta, \subst{x \assign y}A_1 
\\
\multicolumn{3}{r}
{\qquad\qquad
 \mbox{provided $y$ is not free in 
       $\vGamma \union \vDelta \union \set{A_1}$}\;,}
\\
\CExists{x}{A_1}, \vGamma \LCon \vDelta & \mathrm{iff} &
\subst{x \assign y}A_1, \vGamma \LCon \vDelta 
\\
\multicolumn{3}{r}
{\qquad\qquad
 \mbox{provided $y$ is not free in 
       $\vGamma \union \vDelta \union \set{A_1}$}\;,}
\\
\Gamma \LCon \Delta, \Not \False & \!\!\!\!,
\\
\Not (\Not A_1), \Gamma \LCon \Delta         & \mathrm{iff} &
A_1, \Gamma \LCon \Delta\;,
\\
\Not (A_1 \CAnd A_2), \Gamma \LCon \Delta    & \mathrm{iff} &
\Not A_1, \Gamma \LCon \Delta \;\mathrm{and}\;
\Not A_2, \Gamma \LCon \Delta\;,
\\
\Gamma \LCon \Delta, \Not (A_1 \COr A_2)     & \mathrm{iff} &
\Gamma \LCon \Delta, \Not A_1 \;\mathrm{and}\;
\Gamma \LCon \Delta, \Not A_2\;,
\\
\Not (A_1 \IImpl A_2), \Gamma \LCon \Delta   & \mathrm{iff} & 
{A_1, \Not A_2, \Gamma \LCon \Delta\;,}
\\
\Not \CForall{x}{A_1}, \vGamma \LCon \vDelta & \mathrm{iff} &
\Not \subst{x \assign y}A_1, \vGamma \LCon \vDelta 
\\
\multicolumn{3}{r}
{\qquad\qquad
 \mbox{provided $y$ is not free in 
       $\vGamma \union \vDelta \union \set{A_1}$}\;,}
\\
\vGamma \LCon \vDelta, \Not \CExists{x}{A_1} & \mathrm{iff} &
\vGamma \LCon \vDelta, \Not \subst{x \assign y}A_1 
\\
\multicolumn{3}{r}
{\qquad\qquad
 \mbox{provided $y$ is not free in 
       $\vGamma \union \vDelta \union \set{A_1}$}\;,}
\end{array}
\]
\[
\renewcommand{\arraystretch}{1.25}
\begin{array}[t]{r@{\;\;}c@{\;\;}l}
\LCon t_1 = t_1 & \mathrm{and} &
t_1 = t_2, \subst{x \assign t_1}A_1 \LCon \subst{x \assign t_2}A_1
\end{array}
\]
and the logical consequence relation $\clLCon$ of \FOCL\ is the smallest 
logical consequence relation that satisfies the same conditions as 
above, except that $\LCon$ and $\notLCon$ are replaced by $\clLCon$ and 
$\notclLCon$, respectively, and in addition the following conditions:
\[
\renewcommand{\arraystretch}{1.25}
\begin{array}[t]{r@{\;\;}c@{\;\;}l}
\Not A_1, A_1 \clLCon \False                   & \mathrm{and} &
\Not \False \clLCon A_1, \Not A_1\;.
\end{array}
\]
\end{proposition}

The most inartificial paradefinite logic relative to \FOCL\ is the logic 
that differs from \FOCL\ only in that it lacks exactly those classical 
logical consequences that exist due to either LNC or LEM.
This makes that \foBDif\ deserves to be qualified as the most natural 
paradefinite logic relative to \FOCL.

\section{Validity, Satisfiability, and Embedding into $\FOCL$}
\label{EMBEDDING}

This section concerns validity, satisfiability, and the relationship 
between the two in the setting of \foBDif\ as well as a simple embedding 
of $\foBDif$ into $\FOCL$. \linebreak[2]
Validity, satisfiability, and the relationship between them are central 
in automated theorem proving and machine learning.
Moreover, in \foBDif, as in \FOCL, validity and satisfiability can be 
defined in terms of logical consequence.
Therefore the embedding is not only theoretically interesting, but also 
potentially practically relevant.

\subsection{Validity and Satisfiability}

Validity of formulas and satisfiability of sets of formulas are semantic 
notions that are generally considered relevant to any logic.
As in classical logic, these notions are closely related in $\foBDif$.

\begin{sdef}
\label{def-val-sat}
Let $A \in \SForm$ and $\Gamma \subseteq \SForm$.
Then validity of $A$ and satisfiability of $\Gamma$ are defined in terms 
of the logical consequence relation $\LCon$ of $\foBDif$ as follows:
\vspace*{-1.125ex} \par
\[
\renewcommand{\arraystretch}{1.125}
\begin{tabular}[t]{rl}
$A$ is \emph{valid} & iff $\LCon A$;
\\
$\Gamma$ is \emph{satisfiable} & iff $\Gamma \notLCon \False$.
\end{tabular}
\]
$A$ is \emph{invalid} iff $A$ is not valid. 
$\Gamma$ is \emph{unsatisfiable} iff $\Gamma$ is not satisfiable.
\end{sdef}

The following corollary of Definition~\ref{def-val-sat} provides a 
justification of the definition of satisfiability in terms of the 
logical consequence relation.
\begin{corollary}
\label{corollary-just-sat}
Let $\Gamma \subseteq \SForm$.
Then $\Gamma$ is satisfiable iff 
there exists a structure $\mathbf{A}$ of $\foBDif$ and an assignment 
$\alpha$ in $\mathbf{A}$ such that
$\Term{A}{\mathbf{A}}{\alpha} \in \nolinebreak \DValue$  for all 
$A \in \vGamma$.
\end{corollary}

The way satisfiability and validity are related in \foBDif\ is very 
similar to the way in which they are related in classical logic.
The following proposition follows easily from the definitions of 
validity and satisfiability and the valuation of formulas of the 
form $A \IImpl \False$.
\begin{proposition}
\label{prop-valid-sat}
For all $A \in \SForm$:
\vspace*{-1.125ex} \par
\[
\renewcommand{\arraystretch}{1.125}
\begin{tabular}[t]{rl}
$A$ is valid & iff $A \IImpl \False$ is unsatisfiable;
\\
$A$ is satisfiable & iff $A \IImpl \False$ is invalid.
\end{tabular}
\]
\end{proposition}
$A \IImpl \False$ is logically equivalent to $\Not A$ in \FOCL.
However, Proposition~\ref{prop-valid-sat} does not hold if 
$A \IImpl \False$ is replaced by $\Not A$.

The way in which inconsistency and unsatisfiability are related in 
\foBDif\ differ from the way in which they are related in classical 
logic.
\begin{sdef}
\label{def-incons}
Let $\Gamma \subseteq \SForm$.
Then $\Gamma$ is \emph{inconsistent} iff there exists an 
$A \in \nolinebreak \SForm$ such that 
$\Gamma \LCon A$ and $\Gamma \LCon  \Not A$.
$\Gamma$ is \emph{consistent} iff $\Gamma$ is not inconsistent.
\end{sdef}
As a corollary of Definitions~\ref{def-val-sat} and~\ref{def-incons}, we
have that in \foBDif, unlike in \FOCL, inconsistency is not a sufficient 
condition for unsatisfiability.
\begin{corollary}
\label{corollary-incons-sat}
There exists a $\Gamma \subseteq \SForm$ such that $\Gamma$ is both
inconsistent and satisfiable.
\end{corollary}
As a matter of fact, each $\Gamma \subseteq \SForm$ in which the 
connective $\False$ does not occur is satisfiable. 

\subsection{Embedding of $\foBDif$ into $\FOCL$}
\label{subsect-embed}

Because of the classical nature of the connectives and quantifiers of 
$\foBDif$, there exists a simple embedding of $\foBDif$ into $\FOCL$.
 
The embedding concerned is potentially practically relevant.
To give an example, the embedding can be useful to determine, 
for a fragment for which the validity of formulas is known to be 
decidable in $\FOCL$, whether the validity of formulas is decidable in 
$\foBDif$ and to adapt, for such a fragment, an existing decision 
procedure for the validity of formulas in $\FOCL$ to the validity of 
formulas in $\foBDif$.

The embedding is a function from the set of all formulas of a fixed but 
arbitrary instance of $\foBDif$ to the set of all formulas of an 
instance of $\FOCL$ that preserves logical consequence.
The alphabet of the instance of $\FOCL$ differs from the alphabet of the 
instance of $\foBDif$ in that, for each $n \in \Nat$, the set 
$\Pred{n}'$ of all predicate symbols of the instance of $\FOCL$ is
$\Pred{n}$ extended as follows: 
$\Pred{n}' = \Pred{n} \union \set{\denial{P} \where P \in \Pred{n}}$.

\begin{sdef}
\label{def-embedding}
The embedding, denoted by $\Embed{\ph}{}{}$, is inductively defined in 
Table~\ref{table-embedding}. 
\begin{table}[!t]
\caption{Embedding of $\foBDif$ into $\FOCL$}
\label{table-embedding}
\vspace*{-3ex} \par \mbox{} \centering
\renewcommand{\arraystretch}{1.275}
\begin{array}[t]{rcl}
\hline
\mbox{} \\[-2.5ex]
\Embed{p}{}{} & = & \Tp\;,
\\
\Embed{P(t\sb1 ,\ldots, t\sb{n+1})}{}{} & = &
  \TP(t\sb1 ,\ldots, t\sb{n+1})\;,
\\
\Embed{t\sb1 = t\sb2}{}{} & = & t\sb1 \XEq t\sb2\;,
\\
\Embed{\False}{}{} & = & \False \;,
\\
\Embed{A\sb1 \CAnd A\sb2}{}{} & = &
  \Embed{A\sb1}{}{} \CAnd \Embed{A\sb2}{}{} \;,
\\
\Embed{A\sb1 \COr A\sb2}{}{} & = &
  \Embed{A\sb1}{}{} \COr \Embed{A\sb2}{}{} \;,
\\
\Embed{A\sb1 \IImpl A\sb2}{}{} & = &
  \Embed{A\sb1}{}{} \IImpl \Embed{A\sb2}{}{} \;,
\\
\Embed{\CForall{x}{A}}{}{} & = &
  \CForall{x}{\Embed{A}{}{}} \;,
\\
\Embed{\CExists{x}{A}}{}{} & = &
  \CExists{x}{\Embed{A}{}{}} \;,
\\[1ex]
\Embed{\Not p}{}{} & = & \denial{\Tp}\;,
\\
\Embed{\Not P(t\sb1 ,\ldots, t\sb{n+1})}{}{} & = &
  \denial{\TP}(t\sb1 ,\ldots, t\sb{n+1})\;,
\\
\Embed{\Not\; t\sb1 = t\sb2}{}{} & = &
  t\sb1 \DXEq t\sb2\;,
\\
\Embed{\Not \False}{}{} & = & \Not \False \;,
\\
\Embed{\Not \Not A}{}{} & = & \Embed{A}{}{} \;,
\\
\Embed{\Not (A\sb1 \CAnd A\sb2)}{}{} & = &
  \Embed{\Not A\sb1 \COr \Not A\sb2}{}{} \;,
\\
\Embed{\Not (A\sb1 \COr A\sb2)}{}{} & = &
  \Embed{\Not A\sb1 \CAnd \Not A\sb2}{}{} \;,
\\
\Embed{\Not (A\sb1 \IImpl A\sb2)}{}{} & = &
  \Embed{A\sb1 \CAnd \Not A\sb2}{}{} \;,
\\
\Embed{\Not\, \CForall{x}{A}}{}{} & = &
  \Embed{\CExists{x}{\Not A}}{}{} \;,
\\
\Embed{\Not\, \CExists{x}{A}}{}{} & = &
  \Embed{\CForall{x}{\Not A}}{}{} \;.
\vspace*{1.25ex} \\
\hline
\end{array}
\vspace*{-1.5ex} \par
\end{table}
In this table, 
$x$ ranges over all variables from $\SVar$,
$p$ ranges over all predicate symbols from $\Pred{0}$,
$P$ ranges over all predicate symbols from $\Pred{n+1}$, 
$t_1$, \ldots, $t_{n+1}$ range over all terms from $\STerm$, and 
$A_1$, $A_2$, and $A$ range over all formulas from $\SForm$.
\end{sdef}
The intuition is that $\Embed{A}{}{}$ is a classical-logic formula
stating that the formula $A$ is either true only or both true and false 
in $\foBDif$.

\begin{snot}
We write $\denial{A}$, where $A \in \SAForm$, for $\Embed{\Not A}{}{}$.
\end{snot}

\begin{snot}
We write $\Embed{\Gamma}{}{}$, where $\Gamma \subseteq \SForm$, for
$\set{\Embed{A}{}{} \where A \in \Gamma}$.
\end{snot}

The function $\Embed{\ph}{}{}$ is indeed an embedding of $\foBDif$ into 
$\FOCL$.
\begin{theorem}
\label{theorem-embed}
For all $\Gamma,\Delta \subseteq \SForm$,
$\Gamma \LCon \Delta \;\; \mathrm{iff} \;\;
 \Embed{\Gamma}{}{} \clLCon \Embed{\Delta}{}{}$.
\end{theorem}
\begin{proof}
See Appendix~\ref{EMBED}.
\qed
\end{proof}

\begin{sdef}
\label{def-weak-nnf}
Let $A \in \SForm$. 
Then $A$ is in \emph{weak negation normal form} if each occurrences of 
the connective $\Not$ in $A$ is in a subformula of the form $\Not A'$ 
where $A' \in \SAForm$.
\end{sdef}
Theorem~\ref{theorem-embed} shows indirectly how close $\foBDif$ and 
$\FOCL$ are to each other.
Every formula of $\foBDif$ has a weak negation normal form.
For a formula in weak negation normal form, the given embedding causes 
only minor changes. 
It consists solely of replacing each subformula of the form $\Not A$ by 
$\denial{A}$.
Since a weak negation normal form can be obtained in polynomial time, 
Theorem~\ref{theorem-embed} also shows indirectly that validity of 
formulas in $\foBDif$ is polynomially reducible to validity of formulas 
in $\FOCL$.

The following is a corollary of Theorem~\ref{theorem-embed} concerning 
the relation between validity and satisfiability in \foBDif\ and \FOCL.
\begin{corollary}
\label{corollary-embed}
For all $A \in \SForm$:
\[
\renewcommand{\arraystretch}{1.125}
\begin{tabular}[t]{rl}
$A$ is valid       & iff $\Embed{A}{}{}$ is classically valid;
\\
$A$ is satisfiable & iff $\Embed{A}{}{}$ is classically satisfiable.
\end{tabular}
\]
\end{corollary}

The function $\Embed{\ph}{}{}$ extends from formulas to inference rules 
in the obvious way.
The following is another corollary of Theorem~\ref{theorem-embed}.
\begin{corollary}
\label{corollary-embedding-rules}
Let $R$ be an inference rule of the proof system of $\foBDif$ presented 
in Section~\ref{PROOF-SYSTEM}.
Then $\Embed{R}{}{}$ is a derived inference rules of the proof system of 
$\FOCL$ described in Section~\ref{PROOF-SYSTEM}.
\end{corollary}

Seeing that 
$\Embed{\Not (A\sb1 \IImpl A\sb2)}{}{} =
 \Embed{A\sb1 \CAnd \Not A\sb2}{}{}$, 
one might at first sight doubt whether $\Embed{\ph}{}{}$ is indeed an 
embedding of $\foBDif$ into $\FOCL$.
After all, $A_1 \IImpl A_2 \LEqv \Not A_1 \COr A_2$ does not hold for 
the logical equivalence relation of \foBDif.  
However, as for formulas of the form $\Not (A\sb1 \IImpl A\sb2)$, the 
fact that $\LCon \Not (A_1 \IImpl A_2) \IImpl \Not(\Not A_1 \COr A_2)$ 
and $\LCon \Not(\Not A_1 \COr A_2) \IImpl \Not (A_1 \IImpl A_2)$ hold 
for the logical consequence relation of \foBDif, is sufficient for 
$\Embed{\ph}{}{}$ to be an embedding of $\foBDif$ into $\FOCL$. 

\section{Potentials of \foBDif\ in Inductive Machine Learning}
\label{MACHINE-LEARNING}

This section discusses the potentials of \foBDif\ for dealing with 
inconsistencies and incompletenesses in inductive machine learning.

Machine learning is an important aspect of many artificial intelligence 
applications.
One of the foundational approaches to machine learning is inductive 
learning, which is learning by inferring generalizations of a series of 
specific examples.
Concept learning from positive and negative examples of a concept takes 
a central position in inductive learning.
This section is confined to concept learning.

There are various ways to represent concepts and examples of a concept 
and to define the relation between representations of concepts and 
representations of examples that represents the relation between 
concepts and examples by which examples are related to the concept or 
concepts of which they are examples.
Theoretically interesting settings are settings in which concepts and 
examples are represented by sets of formulas and formulas, respectively, 
of a fragment of first-order logic and the relation between concepts and 
examples by which examples are related to the concept or concepts of 
which they are examples is represented by the logical consequence 
relation of first-order logic restricted to the fragment in question.
Learning in such a setting is known as learning from entailment.%
\footnote
{Logical consequence relations are also sometimes called entailment
 relations.}

The problem of concept learning from entailment is described below in 
the general setting where examples, hypotheses, and background knowledge 
elements may be any formula of $\FOCL$.
Usually, at least some restrictions are imposed on the formulas allowed 
as examples, hypotheses, and background knowledge elements.
\begin{snot}
We write $\denial{\vGamma}$, where $\vGamma \subseteq \SForm$, for the 
set $\set{\Not A \where A \in \vGamma}$.
\end{snot}
\begin{sdef}
\label{def-learning-problem}
The \emph{problem of concept learning from entailment} is the 
following problem:
\begin{itemize}
\item[]
Given:
\begin{itemize}
\item
a set $\calE^+ \subseteq \SForm$ of \emph{positive examples}; 
\item
a set $\calE^- \subseteq \SForm$ of \emph{negative examples}; 
\item
a set $\calB \subseteq \SForm$ of 
\emph{background knowledge elements}
\end{itemize}
\hspace*{.5em} such that
\begin{itemize}
\item
for all $e \in \calE^+$,\, $\calB \notclLCon e$;
\item
$\calB, \denial{\calE^-} \notclLCon \False$.
\end{itemize}
\item[]
Find: 
\begin{itemize}
\item
a set $\calH \subseteq \SForm$ of \emph{hypotheses} 
\end{itemize}
\hspace*{.5em} such that
\begin{itemize}
\item
for all $e \in \calE^+$,\, $\calB, \calH \clLCon e$;
\item
$\calB, \calH, \denial{\calE^-} \notclLCon \False$.
\end{itemize}
\end{itemize}
\end{sdef}
In the above definition, a found set $\calH$ of hypotheses is the 
representation of a concept with the examples represented by the 
elements of $\calE^+$ and the examples represented by the elements of 
$\calE^-$ as positive examples and negative examples, respectively.

In the case of the problem of concept learning from entailment as 
defined above, a solution $\calH$ does not always exist.
It follows easily from the definition that the following is a necessary 
condition for the existence of a solution:
\[
\calB, \calE^+, \denial{\calE^-} \notclLCon \False\;.%
\footnotemark
\]
\footnotetext
{It can be shown by a cardinality argument that this condition is not a 
 sufficient condition for the existence of a solution.}%
In other words, a solution exists only if $\calE^-$ is consistent with 
respect to $\calB \union \calE^+$.
However, in many artificial intelligence applications in which concept
learning is involved, it cannot be ruled out that $\calE^-$ is 
inconsistent with respect to $\calB \union \calE^+$.
The proposed ways to deal with such an inconsistency usually do not have 
their origins in logic. 
They amount to removing the inconsistency by adapting $\calE^+$ and/or 
$\calE^-$ such that they are minimally adapted from a particular point 
of view (see e.g.~\cite{PF18a,MM20a}). 
A notable exception is the way proposed in Chapter~7 of~\cite{Rae91a}.
We return to this exception below.

Another way to deal with inconsistency of $\calE^-$ with respect to 
$\calB \union \calE^+$ is to simply accept it and to learn from the 
logical consequence relation of \foBDif.
\begin{sdef}
\label{def-learning-problem-paradef}
The \emph{problem of concept learning from paradefinite entailment} 
is defined as the problem of concept learning from entailment except 
that $\clLCon$ and $\notclLCon$ are replaced by $\LCon$ 
and $\notLCon$, respectively.
\end{sdef}
It follows easily from Definition~\ref{def-learning-problem-paradef} 
that in the case of the problem of concept learning 
from paradefinite entailment that the following is a necessary condition 
for the existence of a solution:
\[
\calB, \calE^+, \denial{\calE^-} \notLCon \False\;.
\]
This condition is always met if the connective $\False$ does not occur 
in $\calB$, $\calE^+$, and~$\denial{\calE^-}$.

For each $A \in \SForm$, $\Embed{A}{}{}$ can be obtained from $A$ in 
polynomial time (see the remark immediately following 
Definition~\ref{def-weak-nnf}). 
From this and Theorem~\ref{theorem-embed}, it follows directly that the 
problem of learning concepts from paradefinite entailment is 
polynomially reducible to the problem of learning concepts from 
entailment.
In essence, the way to deal with inconsistency proposed in Chapter~7 
of~\cite{Rae91a} implicitly uses this reduction.

In practice, the problem of learning concepts from paradefinite 
entailment has to be restricted to formulas from a fragment of \foBDif\ 
for which satisfiability is decidable and which is further closed under 
$\Not$, $\CAnd$, and $\COr$.
By closure under $\Not$, validity is also decidable; and by closure 
under $\Not$, $\CAnd$, and $\COr$ it is also decidable whether the 
logical consequence relation holds between two finite sets of formulas.
Fragments of \foBDif\ for which satisfiability is decidable and which 
are further closed under $\Not$, $\CAnd$, and $\COr$ include
the two-variable fragment~\cite{Mor75a,GKV97a},
the guarded fragment~\cite{Gra99a}, 
the triguarded fragment~\cite{RS18a}, 
the ordered fragment~\cite{Her90a}, 
the forward fragment~\cite{Bed21a}, and
the adjacent fragment~\cite{BKP23a}.
If membership of such a fragment is preserved by the embedding 
$\Embed{\ph}{}{}$, then the fragment is also practically 
relevant to the problem of learning concepts from paradefinite 
entailment.
It is easy to see that the membership of each of the above-mentioned 
fragments is preserved  by the embedding $\Embed{\ph}{}{}$.

\section{Concluding Remarks}
\label{CONCLUSIONS}

This paper provides a fairly comprehensive overview of \foBDif, an 
expansion of first-order Belnap-Dunn logic whose connectives and 
quantifiers all have a counterpart in classical logic.
The language and logical consequence relation of \foBDif\ have been 
rigorously defined, a sequent calculus proof system for \foBDif\ has 
been presented, and the soundness and completeness of the presented 
proof system have been established.
The propositional fragment of \foBDif\ has already been discussed in 
several earlier papers, including~\cite{AA96a,AA98a,AA17a,Pyn99a}, but 
without exception quite casually.

The close relationship between the logical consequence relations of 
\foBDif\ and \FOCL, the version of classical logic with the same 
language, is illustrated by the minor differences between the presented 
proof system for \foBDif\ and a sound and complete proof system for 
\FOCL.

A clear characterization of the classical nature of the connectives and 
quantifiers of \foBDif\ has been given by means of fifteen classical 
laws of logical equivalence.
These laws distinguish \foBDif\ from the many other logics that are 
usually considered equally classical.

It has been argued that \foBDif\ is the most natural paradefinite logic 
with respect to \FOCL.
This has been done on the basis of the way in which the logical 
consequence relation of \foBDif\ can be obtained from the logical 
consequence relation of \FOCL. 

A simple embedding of \foBDif\ into \FOCL\ has been presented. 
This embedding is simple due to the classical nature of the connectives 
and quantifiers of \foBDif.
The potential of \foBDif\ for dealing with inconsistencies and 
incompletenesses in inductive machine learning has been briefly 
discussed.
This discussion suggest that the presented embedding is not only 
theoretically interesting, but also potentially practically relevant.

The expansions of BD to which \foBDif\ is most closely related are 
BD$\mathrm{\Delta}$~\cite{SO14a}, F4CC~\cite{KZ20a}, and 
QLET$_\mathrm{F}$~\cite{ARCC22a}.  
The main similarities and dissimilarities between \foBDif\ and these
logics are:
\begin{itemize}
\item
\foBDif\ is an expansion of BD whose additional connectives have
a counterpart in classical logic, but BD$\mathrm{\Delta}$, F4CC, and 
QLET$_\mathrm{F}$ are expansions of BD whose additional connectives  
have no counterpart in classical logic;
\item
as far as the propositional fragments of the logics are concerned,
\foBDif\ and BD$\mathrm{\Delta}$ are interdefinable, QLET$_\mathrm{F}$
is definable in \foBDif\ but not vice versa, and \foBDif\ is definable 
in F4CC but not vice versa (see~\cite{Mid24b});
\item
unlike the logical consequence relation of \foBDif, the logical 
consequence relations of BD$\mathrm{\Delta}$ and QLET$_\mathrm{F}$ are 
restricted to sentences (formulas without free variables);
\item
unlike in \foBDif, function symbols of positive arity are excluded 
in BD$\mathrm{\Delta}$ and QLET$_\mathrm{F}$, and all function symbols 
are excluded in F4CC.
\end{itemize}

\bibliographystyle{splncs04}
\bibliography{PCL}

\appendix

\section{Proof of Theorem~\ref{theorem-sound-complete-BD}}
\label{SOUND-COMPLETE}

This appendix contains a proof of Theorem~\ref{theorem-sound-complete-BD}.
This theorem states that, for all $\Gamma,\Delta \subseteq \SForm$,\, 
$\Gamma \LDer \Delta$ iff $\Gamma \LCon \Delta$.
In this appendix, the theorem is split up into a theorem concerning the 
only if part (Theorem~\ref{theorem-sound}) and a theorem concerning the 
if part (Theorem~\ref{theorem-complete}).
The proof of Theorem~\ref{theorem-sound-complete-BD} simply becomes:
\begin{proof}
Theorem~\ref{theorem-sound-complete-BD} is an immediate corollary of 
Theorems~\ref{theorem-sound} and~\ref{theorem-complete} presented below.
\qed
\end{proof}

\begin{theorem}
\label{theorem-sound}
Let $\Gamma$ and $\Delta$ be sets of formulas from $\SForm$. 
Then $\Gamma \LDer \Delta$ only if $\Gamma \LCon \Delta$.
\end{theorem}
\begin{proof}
We first consider the special case that $\Gamma$ and $\Delta$ are 
finite. 
In this case, we have to show that $\Gamma \scEnt \Delta$ is provable 
only if $\Gamma \LCon \Delta$.
This is straightforwardly proved by induction on the length of the proof 
of $\Gamma \scEnt \Delta$.
Now, we can easily prove the general case. 
Assume that there exist finite $\Gamma' \subseteq \Gamma$ and 
$\Delta' \subseteq \Delta$ such that $\Gamma' \scEnt \Delta'$ is 
provable.
Then, by the result about the finite case, $\Gamma' \LCon \Delta'$.
From this and the definition of $\LCon$, it follows that also 
$\Gamma \LCon \Delta$. 
\qed
\end{proof}

Let $\Gamma$ and $\Delta$ be sets of formulas from $\SForm$.
Then $\Delta$ is \emph{cut-free derivable} from $\Gamma$, written 
$\Gamma \LDercf \Delta$, iff there exist finite sets 
$\Gamma' \subseteq \Gamma$ and $\Delta' \subseteq \Delta$ such that the 
sequent $\Gamma' \scEnt \Delta'$ is provable without the inference rule 
Cut.

Theorem~\ref{theorem-complete} states that, for all 
$\Gamma,\Delta \subseteq \SForm$,\, 
$\Gamma \LDercf \Delta$ if $\Gamma \LCon \Delta$.
This means that Theorem~\ref{theorem-complete} is stronger than needed 
for Theorem~\ref{theorem-sound-complete-BD} to be an immediate corollary of 
Theorems~\ref{theorem-sound} and~\ref{theorem-complete}.
An additional immediate corollary of Theorems~\ref{theorem-sound} 
and~\ref{theorem-complete} is a cut-elimination result.

In the proof of Theorem~\ref{theorem-complete} given below, use is made
of two lemmas, one about sets of formulas that are regular and the other 
about the canonical model of such sets of formulas.

Let $\Gamma \subseteq \SForm$.
Then $\Gamma$ is \emph{regular} iff for all $A,A_1,A_2 \in \SForm$ and
$x \in \SVar$:
\begin{itemize}
\setlength{\itemsep}{.5ex}
\item
$\Gamma \not\LDercf A'$ for some $A' \in \SForm$;
\item
$\Gamma \LDercf A$ only if $A \in \Gamma$;
\item
$A_1 \COr A_2 \in \Gamma$ only if 
$A_1 \in \Gamma$ or $A_2 \in \Gamma$;
\item
$A_1 \IImpl A_2 \in \Gamma$ only if 
$A_1 \not\in \Gamma$ or $A_2 \in \Gamma$;
\item
$\CForall{x}{A} \in \Gamma$ iff $\subst{x \assign c}A \in \Gamma$
for all $c \in \Func{0}$;
\item
$\CExists{x}{A} \in \Gamma$ iff $\subst{x \assign c}A \in \Gamma$
for some $c \in \Func{0}$.
\end{itemize}

\begin{lemma}
\label{lemma-regular-ext} 
Let $\Gamma,\Delta \subseteq \SForm$ be such that 
$\Gamma \not\LDercf \Delta$.
Then there exist $\Gamma^+,\Delta^+ \subseteq \SForm$ with
$\Gamma \subseteq \Gamma^+$ and $\Delta \subseteq \Delta^+$ such that
$\Gamma^+ \not\LDercf \Delta^+$, 
$A \in \Gamma^+$ or $A \in \Delta^+$ for all $A \in \SForm$, and
$\Gamma^+$ is regular.
\end{lemma}
\begin{proof}
Expand the language of \foBDif\ by adding a countably infinite set $C$ 
of fresh constant symbols to $\Func{0}$.
Let $\langle c_n \rangle_{n \in \Nat}$ be an enumeration of all constant
symbols in $C$.
Let $\langle A_n \rangle_{n \in \Nat}$ be an enumeration of all formulas
of the expanded language.
Define inductively the sequence 
$\langle (\Gamma_n, \Delta_n) \rangle_{n \in \Nat}$ as follows:
\begin{itemize}
\item
$\Gamma_0 = \Gamma$ and $\Delta_0 = \Delta$;
\item
if $\Gamma_n, A_n \not\LDercf \Delta_n$, then:
\vspace{\itemsep}
\begin{itemize}
\item
if $A_n \equiv \CExists{x}{A}$, then
$\Gamma_{n+1} = \Gamma_n \union \set{A_n,\subst{x \assign c}A}$ and
$\Delta_{n+1} = \Delta_n$, where $c$ is the first constant symbol in the
enumeration of $C$ that does not occur in $\Gamma_n$, $\Delta_n$, and 
$A_n$;
\item
if $A_n \not\equiv \CExists{x}{A}$, then
$\Gamma_{n+1} = \Gamma_n \union \set{A_n}$ and 
$\Delta_{n+1} = \Delta_n$;
\end{itemize}
\item
if $\Gamma_n, A_n \LDercf \Delta_n$, then:
\vspace{\itemsep}
\begin{itemize}
\item
if $A_n \equiv \CForall{x}{A}$, then
$\Gamma_{n+1} = \Gamma_n$ and
$\Delta_{n+1} = \Delta_n \union \set{A_n,\subst{x \assign c}A}$, where 
$c$ is the first constant symbol in the enumeration of $C$ that does not 
occur in $\Gamma_n$, $\Delta_n$, and $A_n$;
\item
if $A_n \not\equiv \CForall{x}{A}$, then
$\Gamma_{n+1} = \Gamma_n$ and 
$\Delta_{n+1} = \Delta_n \union \set{A_n}$.
\end{itemize}
\end{itemize}
Define $\Gamma^+$ and $\Delta^+$ as follows:
$\Gamma^+ = \Union_{n \in \Nat} \Gamma_n$ and
$\Delta^+ = \Union_{n \in \Nat} \Delta_n$.

By the construction of $\Gamma^+$ and $\Delta^+$,
$\Gamma \subseteq \Gamma^+$ and $\Delta \subseteq \Delta^+$,
$\Gamma^+ \not\LDercf \Delta^+$, and
$A \in \Gamma^+$ or $A \in \Delta^+$ for all $A \in \SForm$.

It is not hard to prove that $\Gamma^+$ is regular.
Here, we only show that $\CForall{x}{A} \in \Gamma^+$ iff 
$\subst{x \assign c}A \in \Gamma^+$ for all $c \in \Func{0}$.
The only if part follows directly from the construction of $\Gamma^+$.
For the if part, we prove the contrapositive, i.e.\
$\CForall{x}{A} \notin \Gamma^+$ only if 
$\subst{x \assign c}A \notin \Gamma^+$ for some $c \in \Func{0}$.
Assume $\CForall{x}{A} \notin \Gamma^+$.
Then, because $\CForall{x}{A} \in \Gamma^+$ or 
$\CForall{x}{A} \in \Delta^+$, we have 
$\CForall{x}{A}\in \Delta^+$.
From this, it follows by the construction of $\Delta^+$ that 
$\subst{x \assign c}A \in \Delta^+$ for some $c \in \Func{0}$. 
This makes it easy to prove that 
$\subst{x \assign c}A \notin \Gamma^+$ for some $c \in \Func{0}$.
Suppose by contradiction that $\subst{x \assign c}A \in \Gamma^+$ for 
all $c \in \Func{0}$. 
Then, it follows from $\subst{x \assign c}A \in \Delta^+$ for some 
$c \in \Func{0}$ that $\Gamma^+ \LDercf \Delta^+$.
This contradicts the property of $\Gamma^+$ and $\Delta^+$ that 
$\Gamma^+ \not\LDercf \Delta^+$.
Hence, $\CForall{x}{A} \in \Gamma^+$ iff 
$\subst{x \assign c}A \in \Gamma^+$ for all $c \in \Func{0}$.
\qed
\end{proof}

Let $\Gamma \subseteq \SForm$ be regular and let $t_1,t_2 \in \STerm$.
Then $t_1$ is $\Gamma$-\emph{equivalent} to $t_2$, written 
$t_1 \sim_\Gamma t_2$, iff $\Gamma \LDercf t_1 = t_2$.
We write $[t]_\Gamma$, where $t \in \STerm$, for the equivalence class 
of $t$ with respect to $\sim_\Gamma$.

Let $\Gamma \subseteq \SForm$ be regular.
Then the \emph{canonical model} of $\Gamma$, written $\mathbf{A}_\Gamma$,
is the structure of \foBDif\ such that:
\begin{itemize}
\setlength{\itemsep}{.5ex}
\item
$\mathcal{U}^{\mathbf{A}_\Gamma}$ is the set
$\set{[t]_\Gamma \where t \in \STerm}$;
\item
$\mathcal{I}^{\mathbf{A}_\Gamma}(c) = [c]_\Gamma$ 
for every $c \in \Func{0}$;
\item
$\mathcal{I}^{\mathbf{A}_\Gamma}(f)([t_1]_\Gamma,\ldots,[t_n]_\Gamma) =
 [f(t_1,\ldots,t_n)]_\Gamma$
for every $f \in \Func{n+1}$ and $n \in \Nat$; 
\item
$\mathcal{I}^{\mathbf{A}_\Gamma}(p) =
 \left \{
 \begin{array}{@{}l@{\;\;}l@{}}
 \VTrue    & \mathrm{if}\;
 \Gamma \LDercf p \;\mathrm{and}\; \Gamma \not\LDercf \Not p 
 \\
 \VBoth    & \mathrm{if}\;
 \Gamma \LDercf p \;\mathrm{and}\; \Gamma \LDercf \Not p 
 \\
 \VNeither & \mathrm{if}\;
 \Gamma \not\LDercf p \;\mathrm{and}\; \Gamma \not\LDercf \Not p 
 \\
 \VFalse   & \mathrm{if}\;
 \Gamma \not\LDercf p \;\mathrm{and}\; \Gamma \LDercf \Not p 
 \end{array}
 \right.$ 
\\
for every $p \in \Pred{0}$;
\item
$\mathcal{I}^{\mathbf{A}_\Gamma}(P)([t_1]_\Gamma,\ldots,[t_n]_\Gamma) =\!
 \left \{
 \begin{array}{@{}l@{\;\;}l@{}}
 \VTrue    & \mathrm{if}\;
 \Gamma \LDercf P(t_1,\ldots,t_n) \;\mathrm{and}\;
 \Gamma \not\LDercf \Not P(t_1,\ldots,t_n) 
 \\
 \VBoth    & \mathrm{if}\;
 \Gamma \LDercf P(t_1,\ldots,t_n) \;\mathrm{and}\;
 \Gamma \LDercf \Not P(t_1,\ldots,t_n) 
 \\
 \VNeither & \mathrm{if}\;
 \Gamma \not\LDercf P(t_1,\ldots,t_n) \;\mathrm{and}\;
 \Gamma \not\LDercf \Not P(t_1,\ldots,t_n) 
 \\
 \VFalse   & \mathrm{if}\;
 \Gamma \not\LDercf P(t_1,\ldots,t_n) \;\mathrm{and}\;
 \Gamma \LDercf \Not P(t_1,\ldots,t_n) 
 \end{array}
 \right.$ 
\\
for every $P \in \Pred{n+1}$ and $n \in \Nat$.
\end{itemize}

\begin{lemma}
\label{lemma-regular-can} 
Let $\Gamma \subseteq \SForm$ be regular and
let $\alpha$ be the assignment in $\mathbf{A}_\Gamma$ such that
$\alpha(x) = [x]_\Gamma$ for all $x \in \SVar$.
Then, for all $A \in \SForm$:
\[
\displstretch
\begin{array}{l}
\Term{A}{\mathbf{A}_\Gamma}{\alpha} \in \set{\VTrue,\VBoth}
 \;\mathrm{iff}\; A \in \Gamma\;, 
\\
\Term{A}{\mathbf{A}_\Gamma}{\alpha} \in \set{\VFalse,\VBoth}
 \;\mathrm{iff}\; \Not A \in \Gamma\;. 
\end{array}
\] 
\end{lemma}
\begin{proof}
This is proved by induction on the structure of $A$.
Here, we only consider the cases where $A \equiv \Not A'$ and
$A \equiv A'_1 \IImpl A'_2$.

The case $A \equiv \Not A'$:
\begin{itemize}
\setlength{\itemsep}{.5ex}
\item
$\Term{\Not A'}{\mathbf{A}_\Gamma}{\alpha} \in \set{\VTrue,\VBoth}$ iff
$\Term{A'}{\mathbf{A}_\Gamma}{\alpha} \in \set{\VFalse,\VBoth}$ by the 
definition of $\Term{\_\hspace*{.08em}}{\mathbf{A}_\Gamma}{\alpha}$;
\item
$\Term{A'}{\mathbf{A}_\Gamma}{\alpha} \in \set{\VFalse,\VBoth}$ iff
$\Not A' \in \Gamma$ by the induction hypothesis
\end{itemize}
and
\begin{itemize}
\setlength{\itemsep}{.5ex}
\item
$\Term{\Not A'}{\mathbf{A}_\Gamma}{\alpha} \in \set{\VFalse,\VBoth}$ iff
$\Term{A'}{\mathbf{A}_\Gamma}{\alpha} \in \set{\VTrue,\VBoth}$ by the 
definition of $\Term{\_\hspace*{.08em}}{\mathbf{A}_\Gamma}{\alpha}$; 
\item
$\Term{A'}{\mathbf{A}_\Gamma}{\alpha} \in \set{\VTrue,\VBoth}$ iff
$A' \in \Gamma$ by the induction hypothesis; 
\item
$A' \in \Gamma$ iff $\Not (\Not A') \in \Gamma$ by the regularity of 
$\Gamma$.
\end{itemize}

The case $A \equiv A'_1 \IImpl A'_2$:
\begin{itemize}
\setlength{\itemsep}{.5ex}
\item
$\Term{A'_1 \IImpl A'_2}{\mathbf{A}_\Gamma}{\alpha} \in
 \set{\VTrue,\VBoth}$ iff
$\Term{A'_1}{\mathbf{A}_\Gamma}{\alpha} \in \set{\VFalse,\VNeither}$ or
$\Term{A'_2}{\mathbf{A}_\Gamma}{\alpha} \in \set{\VTrue,\VBoth}$ by the 
definition of $\Term{\_\hspace*{.08em}}{\mathbf{A}_\Gamma}{\alpha}$;
\item
$\Term{A'_1}{\mathbf{A}_\Gamma}{\alpha} \in \set{\VFalse,\VNeither}$ or
$\Term{A'_2}{\mathbf{A}_\Gamma}{\alpha} \in \set{\VTrue,\VBoth}$ iff
$A'_1 \notin \Gamma$ or $A'_2 \in \Gamma$ by the induction 
hypothesis;
\item
$A'_1 \notin \Gamma$ or $A'_2 \in \Gamma$ iff 
$A'_1 \IImpl A'_2 \in \Gamma$ by the regularity of $\Gamma$
\end{itemize}
and
\begin{itemize}
\setlength{\itemsep}{.5ex}
\item
$\Term{A'_1 \IImpl A'_2}{\mathbf{A}_\Gamma}{\alpha} \in
 \set{\VFalse,\VBoth}$ iff
$\Term{A'_1}{\mathbf{A}_\Gamma}{\alpha} \in \set{\VTrue,\VBoth}$ and
$\Term{A'_2}{\mathbf{A}_\Gamma}{\alpha} \in \set{\VFalse,\VBoth}$ by the 
definition of $\Term{\_\hspace*{.08em}}{\mathbf{A}_\Gamma}{\alpha}$;
\item
$\Term{A'_1}{\mathbf{A}_\Gamma}{\alpha} \in \set{\VTrue,\VBoth}$ and
$\Term{A'_2}{\mathbf{A}_\Gamma}{\alpha} \in \set{\VFalse,\VBoth}$ iff
$A'_1 \in \Gamma$ and $\Not A'_2 \in \Gamma$ by the \nolinebreak[2] 
induc\-tion hypothesis;
\item
$A'_1 \in \Gamma$ and $\Not A'_2 \in \Gamma$ only if 
$\Not (A'_1 \IImpl A'_2) \in \Gamma$ by the regularity of $\Gamma$;
\item
$A'_1 \notin \Gamma$ or $\Not A'_2 \notin \Gamma$ only if 
$\Not (A'_1 \IImpl A'_2) \notin \Gamma$ --- the contrapositive of 
$A'_1 \in \Gamma$ and $\Not A'_2 \in \Gamma$ if 
$\Not (A'_1 \IImpl A'_2) \in \Gamma$ --- is proved as follows:
\vspace{\itemsep}
\begin{itemize}
\setlength{\itemsep}{.5ex}
\item
$A'_1 \notin \Gamma$ or $\Not A'_2 \notin \Gamma$ iff
$\Term{A'_1}{\mathbf{A}_\Gamma}{\alpha} \in \set{\VFalse,\VNeither}$ or
$\Term{\Not A'_2}{\mathbf{A}_\Gamma}{\alpha} \in
 \set{\VFalse,\VNeither}$ by the \nolinebreak[2] induc\-tion hypothesis;
\item
$\Term{A'_1}{\mathbf{A}_\Gamma}{\alpha} \in \set{\VFalse,\VNeither}$ or
$\Term{\Not A'_2}{\mathbf{A}_\Gamma}{\alpha}
 \in \set{\VFalse,\VNeither}$ iff
$\Term{A'_1}{\mathbf{A}_\Gamma}{\alpha} \in \set{\VFalse,\VNeither}$ or
$\Term{A'_2}{\mathbf{A}_\Gamma}{\alpha} \in \set{\VTrue,\VNeither}$ by
the definition of $\Term{\_\hspace*{.08em}}{\mathbf{A}_\Gamma}{\alpha}$;
\item
the case 
$\Term{A'_1}{\mathbf{A}_\Gamma}{\alpha} \in \set{\VFalse,\VNeither}$: 
\vspace{\itemsep}
\begin{itemize}
\setlength{\itemsep}{.5ex}
\item
$\Term{A'_1}{\mathbf{A}_\Gamma}{\alpha} \in \set{\VFalse,\VNeither}$ 
only if $\Term{A'_1 \IImpl A'_2}{\mathbf{A}_\Gamma}{\alpha} = \VTrue$ by
the definition of $\Term{\_\hspace*{.08em}}{\mathbf{A}_\Gamma}{\alpha}$;
\item
$\Term{A'_1 \IImpl A'_2}{\mathbf{A}_\Gamma}{\alpha} = \VTrue$ iff
$\Term{\Not (A'_1 \IImpl A'_2)}{\mathbf{A}_\Gamma}{\alpha} = \VFalse$ by
the definition of $\Term{\_\hspace*{.08em}}{\mathbf{A}_\Gamma}{\alpha}$;
\item
$\Term{\Not (A'_1 \IImpl A'_2)}{\mathbf{A}_\Gamma}{\alpha} = \VFalse$ 
only if $\Not (A'_1 \IImpl A'_2) \notin \Gamma$ by 
Theorem~\ref{theorem-sound};
\end{itemize}
\item
the case 
$\Term{A'_2}{\mathbf{A}_\Gamma}{\alpha} \in \set{\VTrue,\VNeither}$: 
\vspace{\itemsep}
\begin{itemize}
\setlength{\itemsep}{.5ex}
\item
$\Term{A'_2}{\mathbf{A}_\Gamma}{\alpha} \in \set{\VTrue,\VNeither}$ 
only if 
$\Term{A'_1 \IImpl A'_2}{\mathbf{A}_\Gamma}{\alpha} \in
 \set{\VTrue,\VNeither}$ by the definition of
$\Term{\_\hspace*{.08em}}{\mathbf{A}_\Gamma}{\alpha}$;
\item
$\Term{A'_1 \IImpl A'_2}{\mathbf{A}_\Gamma}{\alpha} \in
 \set{\VTrue,\VNeither}$ iff
$\Term{\Not (A'_1 \IImpl A'_2)}{\mathbf{A}_\Gamma}{\alpha} \in
 \set{\VFalse,\VNeither}$ by the definition of  
$\Term{\_\hspace*{.08em}}{\mathbf{A}_\Gamma}{\alpha}$;
\item
$\Term{\Not (A'_1 \IImpl A'_2)}{\mathbf{A}_\Gamma}{\alpha} \in
 \set{\VFalse,\VNeither}$ 
only if $\Not (A'_1 \IImpl A'_2) \notin \Gamma$ by 
Theorem~\ref{theorem-sound}.
\end{itemize}
\end{itemize}
\end{itemize}
\qed
\end{proof}

\begin{theorem}
\label{theorem-complete}
Let $\Gamma$ and $\Delta$ be sets of formulas from $\SForm$. 
Then $\Gamma \LDercf \Delta$ if $\Gamma \LCon \Delta$.
\end{theorem}
\begin{proof}
We prove the contrapositive, i.e.\ 
$\Gamma \not\LDercf \Delta$ only if $\Gamma \not\LCon \Delta$.
Assume $\Gamma \not\LDercf \Delta$.
Then by Lemma~\ref{lemma-regular-ext}, there exist 
$\Gamma^+,\Delta^+ \subseteq \SForm$ with $\Gamma \subseteq \Gamma^+$ 
and $\Delta \subseteq \Delta^+$ such that 
$\Gamma^+ \not\LDercf \Delta^+$, and $\Gamma^+$ is regular.
Let $\alpha$ be the assignment in $\mathbf{A}_{\Gamma^+}$ such that
$\alpha(x) = [x]_{\Gamma^+}$ for all $x \in \SVar$.
By Lemma~\ref{lemma-regular-can}, 
$\Term{A}{\mathbf{A}_{\Gamma^+}}{\alpha} \in \set{\VTrue,\VBoth}$ for 
all $A \in \Gamma$ and 
$\Term{A'}{\mathbf{A}_{\Gamma^+}}{\alpha} \notin \set{\VTrue,\VBoth}$ 
for all $A' \in \Delta$ (because otherwise we would have 
$\Gamma \LDercf A'$ for some $A' \in \Delta$ which contradicts the 
assumption that $\Gamma \not\LDercf \Delta$).
Hence, $\Gamma \not\LCon \Delta$.
\qed
\end{proof}
The proof of Theorem~\ref{theorem-complete} has been inspired by the 
completeness proof for a natural deduction proof system for 
BD$\mathrm{\Delta}$ in~\cite{SO14a}.
The main differences between the proofs are due to the following: 
(a)~function symbols of positive arity are excluded from the alphabet of 
BD$\mathrm{\Delta}$ and 
(b)~the logical consequence relation of BD$\mathrm{\Delta}$ is 
restricted to sentences (formulas without free variables).

Immediate corollaries of Theorems~\ref{theorem-sound} 
and~\ref{theorem-complete} are Theorem~\ref{theorem-sound-complete-BD} and 
the following cut-elimination result.
\begin{corollary}
Let $\Gamma$ and $\Delta$ be sets of formulas from $\SForm$. 
Then $\Gamma \LDercf \Delta$ if $\Gamma \LDer \Delta$.
\end{corollary}

\section{Proof of Theorem~\ref{theorem-uniqueness}}
\label{UNIQUENESS}

This appendix contains a proof of Theorem~\ref{theorem-uniqueness}.
This theorem states that there is exactly one logic whose matrix is 
four-valued, regular, and classically closed and whose logical 
equivalence relation satisfies laws \mbox{(1)--(15)} from 
Table~\ref{laws-lequiv}.
The proof goes as follows:
\begin{proof}
\sloppy
It follows from the regularity and classical closedness of such a matrix 
that the function $\TFunct(\CAnd)$ is such that:
\begin{cdispl}
\begin{eqncol}
\TFunct(\CAnd)(\VTrue,\VTrue)   = \VTrue\;, \\  
\TFunct(\CAnd)(\VFalse,\VTrue)  = \VFalse\;, \\
\TFunct(\CAnd)(\VTrue,\VFalse)  = \VFalse\;, \\
\TFunct(\CAnd)(\VFalse,\VFalse) = \VFalse\;, 
\end{eqncol}
\qquad
\begin{eqncol}
\TFunct(\CAnd)(\VBoth,\VTrue)       \in \DValue\;,  \\ 
\TFunct(\CAnd)(\VNeither,\VTrue)    \in \NDValue\;, \\ 
\TFunct(\CAnd)(\VTrue,\VBoth)       \in \DValue\;,  \\ 
\TFunct(\CAnd)(\VTrue,\VNeither)    \in \NDValue\;, \\ 
\TFunct(\CAnd)(\VBoth,\VBoth)       \in \DValue\;,  \\ 
\TFunct(\CAnd)(\VNeither,\VNeither) \in \NDValue\;,    
\end{eqncol}
\qquad
\begin{eqncol}
\TFunct(\CAnd)(\VBoth,\VFalse)    \in \NDValue\;, \\ 
\TFunct(\CAnd)(\VFalse,\VBoth)    \in \NDValue\;, \\ 
\TFunct(\CAnd)(\VNeither,\VFalse) \in \NDValue\;, \\ 
\TFunct(\CAnd)(\VFalse,\VNeither) \in \NDValue\;, \\ 
\TFunct(\CAnd)(\VNeither,\VBoth)  \in \NDValue\;, \\ 
\TFunct(\CAnd)(\VBoth,\VNeither)  \in \NDValue\;.    
\end{eqncol}
\end{cdispl}%
So, there are $2^{12}$ alternatives for $\TFunct(\CAnd)$.
However, law~(3) excludes 
$\TFunct(\CAnd)(\VBoth,\VTrue) = \VTrue$ and
$\TFunct(\CAnd)(\VNeither,\VTrue) = \VFalse$,
laws~(3) and~(7) exclude 
$\TFunct(\CAnd)(\VTrue,\VBoth) = \VTrue$ and
$\TFunct(\CAnd)(\VTrue,\VNeither) = \VFalse$,
law~(5) excludes 
$\TFunct(\CAnd)(\VBoth,\VBoth) = \VTrue$ and
$\TFunct(\CAnd)(\VNeither,\VNeither) = \VFalse$, 
law~(1) excludes 
$\TFunct(\CAnd)(\VBoth,\VFalse) = \VNeither$,
$\TFunct(\CAnd)(\VFalse,\VBoth) = \VNeither$,
$\TFunct(\CAnd)(\VNeither,\VFalse) = \VNeither$, and
$\TFunct(\CAnd)(\VFalse,\VNeither) = \VNeither$, and
laws~(9) and~(11), together with the condition imposed on 
$\TFunct(\COr)$ in the case of a regular matrix, exclude
$\TFunct(\CAnd)(\VNeither,\VBoth) = \VNeither$ and
$\TFunct(\CAnd)(\VBoth,\VNeither) = \VNeither$.
Hence, laws~(1), (3), (5), (7), (9), and~(11) exclude all but one 
of the $2^{12}$ alternatives for $\TFunct(\CAnd)$.

Similarly, laws~(2), (4), (6), (8), (10), and~(11) exclude all but one 
of the $2^{12}$ alternatives for $\TFunct(\COr)$,
laws~(12) and~(13)  exclude all but one 
of the $2^{12}$ alternatives for $\TFunct(\IImpl)$, and
law~(11) excludes all but one 
of the $4$ alternatives for $\TFunct(\Not)$.

Moreover, laws~(14) and~(15), together with the remaining alternatives 
for $\TFunct(\CAnd)$ and $\TFunct(\COr)$, exclude all alternatives 
for $\TFunct(\forall)$ and $\TFunct(\exists)$ where, for some 
non-empty $V \subseteq \TValue$, $\TFunct(\forall)(V) \neq \inf V$ 
and $\TFunct(\exists)(V) \neq \sup V$, respectively.
\qed
\end{proof}

\section{Proof of Theorem~\ref{theorem-embed}}
\label{EMBED}

This appendix contains a proof of Theorem~\ref{theorem-embed}.
This theorem states that, for all $\Gamma,\Delta \subseteq \SForm$,
$\Gamma \LCon \Delta \;\; \mathrm{iff} \;\;
 \Embed{\Gamma}{}{} \clLCon \Embed{\Delta}{}{}$.
The proof goes as follows:
\begin{proof}
The only if part is proved as follows.
By Theorems~\ref{theorem-sound-complete-BD} 
and~\ref{theorem-sound-complete-CL}, it is sufficient to prove that,
for all finite $\Gamma,\Delta \subseteq \SForm$,
\[
\begin{array}[t]{@{}l@{}}
\Gamma \scEnt \Delta \;\text{is provable in}\; \foBDif
\quad \text{only if} \quad
\Embed{\Gamma}{}{} \scEnt \Embed{\Delta}{}{}
\;\text{is provable in}\; \FOCL\;.
\end{array}
\]
This is easily proved by induction on the length of a proof 
of $\Gamma \scEnt \Delta$ and case distinction on the last inference 
rule applied, using that the sequent calculus proof system of \FOCL\ 
described in Section~\ref{PROOF-SYSTEM} contains all inference rules of 
$\foBDif$.

The if part is proved by contrapositive.
Let $\mathbf{A}$ be a structure of $\foBDif$.
Then $\mathbf{A}$ can be transformed into a structure $\mathbf{A\sp{*}}$ 
of \smash{$\FOCL$} with the property that, 
for all assignments $\alpha$ in $\mathbf{A}$,
for all atomic formula $A \in \SAForm$:
\[
\renewcommand{\arraystretch}{1.25}
\begin{tabular}[t]{l}
$\Term{A}{\mathbf{A\sp{*}}}{\alpha} = \VTrue$ and
$\Term{\denial{A}}{\mathbf{A\sp{*}}}{\alpha} = \VFalse$ \,iff\,
$\Term{A}{\mathbf{A}}{\alpha} = \VTrue$,
\\
$\Term{A}{\mathbf{A\sp{*}}}{\alpha} = \VTrue$ and
$\Term{\denial{A}}{\mathbf{A\sp{*}}}{\alpha} = \VTrue$ \,iff\,
$\Term{A}{\mathbf{A}}{\alpha} = \VBoth$,
\\
$\Term{A}{\mathbf{A\sp{*}}}{\alpha} = \VFalse$ and
$\Term{\denial{A}}{\mathbf{A\sp{*}}}{\alpha} = \VFalse$ \,iff\,
$\Term{A}{\mathbf{A}}{\alpha} = \VNeither$,
\\
$\Term{A}{\mathbf{A\sp{*}}}{\alpha} = \VFalse$ and
$\Term{\denial{A}}{\mathbf{A\sp{*}}}{\alpha} = \VTrue$ \,iff\,
$\Term{A}{\mathbf{A}}{\alpha} = \VFalse$.
\end{tabular}
\]
Now assume that the structure $\mathbf{A}$ and an assignment $\alpha$ in 
$\mathbf{A}$ form a counter-example for $\Gamma \LCon \Delta$.
Then, it follows straightforwardly from the above men\-tioned property 
of the structure $\mathbf{A\sp{*}}$ that $\mathbf{A\sp{*}}$ and $\alpha$ 
form a counter-example for \linebreak[2]
$\Embed{\Gamma}{}{} \clLCon \Embed{\Delta}{}{}$.
\qed
\end{proof}

\end{document}